\documentclass{article}
\usepackage{amsmath, amssymb}
\usepackage[dvips]{graphicx}
\allowdisplaybreaks
\begin{document}
\title{A small guide to variations in teleparallel gauge theories of
gravity and the Kaniel-Itin model} 
\author{Uwe~Muench, Frank~Gronwald,
Friedrich\,W.~Hehl\thanks{Institute for Theoretical Physics,
University of Cologne, D-50923 K\"oln, Germany}}
\date{}
\maketitle
\pagestyle{myheadings}
\markright{Variations in teleparallel theories}

\begin{abstract}
  Recently Kaniel \& Itin proposed a gravitational model with the wave
  type equation $[\square+\lambda(x)]\vartheta^\alpha=0$ as vacuum
  field equation, where $\vartheta^\alpha$ denotes the coframe of
  spacetime. They found that the viable Yilmaz-Rosen metric is an
  exact solution of the tracefree part of their field equation. This
  model belongs to the \emph{teleparallelism} class of gravitational
  gauge theories. Of decisive importance for the evaluation of the
  Kaniel-Itin model is the question whether the variation of the
  coframe commutes with the Hodge star.  We find a master formula for
  this commutator and rectify some corresponding mistakes in the
  literature. Then we turn to a detailed discussion of the Kaniel-Itin
  model. \emph{file kaniel21.tex, 1998-01-12}
\end{abstract}

Keywords: gravity, gauge theory, teleparallelism, variation, Hodge duals

\section{Introduction}

We were very much surprised when we learned during the 8$^{\text{th}}$
Marcel Grossmann Meeting in Jerusalem \cite{Piran} that Kaniel \& Itin
\cite{kani97} were able to propose a gravitational model which looks
viable at a first sight even if it had neither an Einstein-Hilbert
type of Lagrangian nor the Schwarzschild metric as an exact solution.
Their gravitational potential is represented by a quartet of 1-forms
$\vartheta^{\hat{0}},\vartheta^{\hat{1}},\vartheta^{\hat{2}},
\vartheta^{\hat{3}}$ or, for short, by $\vartheta^\alpha$, which
constitutes the coframe field of spacetime. Their vacuum field
equation is simply the wave equation with an additional `massive'
contribution depending on some scalar field $\lambda(x)$:
\begin{equation}\label{WAVE}
  \left[\square+\lambda(x)\right]\vartheta^\alpha=0\,.\end{equation}
They show that the Yilmaz-Rosen metric \cite{yilmaz58,rose73} solves
the \emph{tracefree} part of \eqref{WAVE} exactly.

Let us be a bit more specific: The Yilmaz-Rosen metric, in isotropic
coordinates, is given by
\begin{equation}\label{YILMAZ} g=e^{-\frac{2m}{r}}\,dt^2-e^{\frac{2m}{r}}\,
  \left(dx^2 +dy^2 +dz^2\right)\;,\end{equation} where
\(r^2:=x^2+y^2+z^2\). If we introduce an orthonormal coframe,
\begin{equation}\label{ORTHONORMAL}
  g=o_{\alpha\beta}\,\vartheta^\alpha\otimes\vartheta^\beta\qquad
  \text{with} \qquad o_{\alpha\beta} = \text{diag}(+1, -1, -1,
  -1)\;,\end{equation} then the following coframe, up to arbitrary
local Lorentz transformations, represents the Yilmaz-Rosen metric:
\begin{equation}\label{COFRAME}
  \vartheta^{\hat{t}}=e^{-\frac{m}{r}}\,dt\;,\quad
  \vartheta^{\hat{x}}=e^{\frac{m}{r}}\,dx \;,\quad
  \vartheta^{\hat{y}}=e^{\frac{m}{r}}\,dy \;,\quad \text{and}\quad
  \vartheta^{\hat{z}}=e^{\frac{m}{r}}\,dz\;. \end{equation} The
tracefree part of \eqref{WAVE} will be determined in Sec.\ 
\ref{sec:44} and turns out to be
\begin{equation}\label{TRACEFREE}\left[
    \square - \frac{1}{4} (e_\beta \rfloor \square
    \vartheta^\beta)\right]\vartheta^\alpha=0\,.\end{equation} The
coframe \eqref{COFRAME} solves the tracefree field equation
\eqref{TRACEFREE} exactly. We have verified this by means of our
computer algebra program {\tt kaniti.exi} displayed in the appendix in
Sec.\ \ref{sec:ca}.

Kaniel \& Itin tried to derive the field equation \eqref{WAVE} from a
suitable Lagrangian. For that purpose they had to assume specifically
that the variation $\delta\vartheta^\alpha$ of the coframe
$\vartheta^\alpha$ commutes with the Hodge star:
$^\star\delta\vartheta^\alpha=\delta\,^\star\vartheta^\alpha$.
However, such a commutativity is only valid for internal Yang-Mills
fields. It is violated for the coframe and the metric. Therefore the
Kaniel-Itin model is based on somewhat shaky foundations.

In the light of the results mentioned so far, the following questions
come to mind: (i) What is the source on the right hand side of the
field equation \eqref{WAVE}? (ii) Can the Yilmaz-Rosen metric also be
adjusted to the trace part of \eqref{WAVE} and, more generally, to a
possible source term on the right hand side of \eqref{WAVE}? (iii) Is
there a consistent variational principle available which would allow
to derive \eqref{WAVE}, including a source term, from a suitable
Lagrangian?  (iv) What is the (geometrical?)  meaning of the
constrained variations of Kaniel \& Itin?

The purpose of this article is to try to answer these questions.
Moreover, along our way, we will discuss some unclear points on the
commutativity of variation and Hodge star which led to some (so far
uncorrected) mistakes in the literature.---

In Sec.\ \ref{sec:model-KI} we provide some background material on how
to derive wave equations of the general type \eqref{WAVE} from
Lagrangians in Maxwell's theory and in theories of other internal
fields. Here and in the following \emph{internal} fields are those
which do not depend on the spacetime geometry (in contrast to
$\vartheta^\alpha$ and $g_{\alpha\beta}$). In this way we are able to
understand how the Lagrangian of Kaniel-Itin comes up in the first
place. But since they identify the \emph{gravitational potential} with
the \emph{coframe}, we run into trouble from the point of view of
finding a suitable Lagrangian.

Any reasonable gauge approach to gravity contains in some way the
gauging of the translation group. The simplest gauge theories of
gravity are teleparallel theories with only the translation group as
gauge group.  They already require the knowledge of how to vary the
Hodge dual of forms. In a teleparallel theory, spacetime can be
described by an \emph{orthonormal} coframe \(\vartheta^\alpha\) as the
only gravitational field variable, which is interpreted as
translational gauge potential, see \cite{gron96}. And this gauge
potential was used by Kaniel-Itin in their model. Accordingly, their
Lagrangian is a special teleparallelism Lagrangian with the additional
postulate of constrained variations. The commutativity of \(\delta\)
and \({}^\star\) is, in general, \emph{not} fulfilled for gauge
theories of {\it external\/} (or spacetime) groups, i.e., for
gravitational gauge theories. In this case it is important to know the
commutator $\delta{}^\star-{}^\star\delta$ of the variation $\delta$
and the Hodge star $^\star$.

Therefore, in Sec.\ \ref{sec:hodge} we will derive the master formula
\eqref{eq:var-hodge} for $\delta{}^\star-{}^\star\delta$. We will
include general variations of the components \(g_{\alpha\beta}\) of
the Riemannian metric $g$ besides those of a (not necessarily
orthonormal) coframe \(\vartheta^\alpha\). If we \emph{insist}, in
accordance with the Kaniel-Itin postulate, on commutativity of
$\delta$ and $^\star$, then the variations $\delta g_{\alpha\beta}$ of
the components of the metric are no longer independent and can be
expressed in terms of the variation $\delta\vartheta^\alpha$ of the
coframe, see \eqref{constraint}.

In Sec.\ \ref{sec:tele} we give a short overview of teleparallelism
theories and the relevant qua\-dra\-tic Lagrangians. We will discuss
the viable set of Lagrangians and display the results in Table 1. We
will show that the KI-Lagrangian, for arbitrary variations, is not
viable. Some errors in the literature (see Schweitzer et al.\ 
\cite{straumann1,straumann2}) are rectified.

In Sec.\ \ref{sec:KI}, we evaluate the model of Kanin \&
Itin \cite{kani97}. The field equation of the constrained variational
principle is the \emph{antisymmetric part} of a wave equation
for $\vartheta^\alpha$, in contrast to the full wave equation as
claimed by Kaniel and Itin.

The Yilmaz-Rosen metric, found by Yilmaz \cite[Eqs.(18) and
(20)]{yilmaz58} in 1958 as a solution in the context of a scalar field
theory of gravitation, also turned out to be a solution of the
bi-metric theory of gravitation of Rosen \cite{rose73}; cf.\ also
\cite{rose74,yilmaz76}.  And, in the Kaniel-Itin model, it solves the
tracefree wave equation.  In Sec.\ \ref{sec:YR-sol} we compare the
Yilmaz-Rosen with the Schwarzschild metric and give, in Sec.\ 
\ref{YR-mot}, a motivation for the emergence of the Yilmaz-Rosen
metric. Finally, we investigate the implications that would arise if
the Yilmaz-Rosen metric is considered to be a solution of the field
equation of Kaniel \& Itin including its trace. In Sec.\ 6 we collect
our arguments.

\section{Prolegomena to the Kaniel-Itin model}\label{sec:model-KI}

\subsection{Maxwell's theory and the wave equation}
The kinetic part of the Lagrangian of a Yang-Mills theory is
conventionally built from the first derivative of the gauge potential
$A$ and the corresponding Hodge dual. For an internal gauge group,
such as for the $U(1)$ or the $SU(2)$, the gauge potential $A$ is
\emph{independent} of the metric $g$ or the coframe $\vartheta^\alpha$
of the underlying spacetime manifold. Then the variation \(\delta\) of
$A$ commutes with the Hodge star operator \({}^\star\). Let us
illustrate this for Maxwell's theory, i.e., for $U(1)$-gauge theory in
Minkowski spacetime.

The Maxwell Lagrangian is given by\footnote{We are using the calculus
  of exterior differential forms, cf.\ \cite{Choquet,Thirring}. Our
  conventions are fixed in \cite{hehl95}.}
\begin{equation} \label{U1Lagrangian}L_{\rm
    Max}=\frac{1}{2}\,dA\wedge\ensuremath{{}^\star} dA\,.
\end{equation} The variation of the 1-form \(A\) is
independent of the variations \(\delta\vartheta^\alpha\) or \(\delta
g_{\alpha\beta}\); furthermore, it commutes with the exterior
derivative, since the variation is defined in this way.  Therefore,
with the \emph{co}derivative $d^\dagger:=-\ensuremath{{}^\star}
d\ensuremath{{}^\star}$, we find
\begin{equation}\delta 
L_{\rm Max}= d(\delta A\wedge\ensuremath{{}^\star} d A)-\delta A\wedge
\ensuremath{{}^\star} 
d^\dagger d A\,.\end{equation}
Thus the vacuum field equation reads: 
\begin{equation} -\ensuremath{{}^\star}
  d^\dagger d A=0\,.\label{eq:MaxEL}\end{equation} Additionally, we
take the Lorentz condition
\begin{equation} \label{Lorentz}
d^\dagger A=0 \end{equation}
as a gauge
condition. Then, introducing the d'Alembertian 
\begin{equation}\square:= d^\dagger d+d
  d^\dagger=-{}^\star d{}^\star d-d{}^\star d{}^\star\,,\end{equation}
the vacuum field equation can be rewritten as
\begin{equation}
- \square\,\ensuremath{{}^\star} A=0\,. \label{eq:wave-max} \end{equation}

One could try to derive (\ref{eq:wave-max}) directly by supplementing
(\ref{U1Lagrangian}) with a suitable Lagrangian. The choice
\begin{equation} L_{\rm Max}^\dagger:=\frac{1}{2}\,d^\dagger A\wedge
  \ensuremath{{}^\star} d^\dagger A\,
\label{eq:lmax-dagger}\end{equation}
looks suggestive. It leads to the field equation 
\begin{equation}\label{Maxadjoint}
 -\ensuremath{{}^\star}
  d\, d^\dagger A=0\,,
\end{equation}which should be compared with \eqref{eq:MaxEL}. Consequently 
the sum of the Lagrangians \eqref{U1Lagrangian} and
\eqref{eq:lmax-dagger}, enriched by a matter Lagrangian, 
\begin{equation}\label{Alt}
  L_{\rm Max}+L_{\rm Max}^\dagger+ L_{\rm mat}
  =\frac{1}{2}\left(dA\wedge\ensuremath{{}^\star} dA +d^\dagger
    A\wedge\ensuremath{{}^\star} d^\dagger A\right)+ L_{\rm mat}
\end{equation}
would yield directly the wave equation:\begin{equation}\label{Altwave}
  \square\,{}^\star A=\frac{\delta L_{\rm mat}}{\delta A}\,.\end{equation}
However, the Lagrangian (\ref{eq:lmax-dagger}) is not gauge-invariant:
For the regauging by means of the arbitrary function \(f\),
\begin{equation} A \longrightarrow A + df\;,
\label{eq:maxdagger-gauge}\end{equation} 
one finds\footnote{Incidentally, as pointed out by Obukhov, the
  Lagrangian \eqref{eq:lmax-dagger} represents an example for a
  Lagrangian of a gauge theory of \(p\)-forms, see \cite{yuri82a,
    yuri82b}. Then, instead of \eqref{eq:maxdagger-gauge}, one has \(A
  \rightarrow A + d^{\dagger}\phi\) as gauge transformation, since
  \(d^{\dagger}d^{\dagger}=0\). And the new ``Lorentz condition'' is
  \(dA=0\).}
\begin{equation}\label{gaugevariance} 
  L_{\rm Max}^\dagger\longrightarrow L_{\rm Max}^\dagger +
  \ensuremath{{}^\star} d^{\dagger} df \wedge \left( d^{\dagger} A +
    \frac{1}{2} d^{\dagger} df \right) \;.
\end{equation} Accordingly, the Lagrangian \eqref{Alt} has to be rejected. 
We can obtain the wave equation \eqref{eq:wave-max} only in the
\emph{special} gauge \eqref{Lorentz}, \emph{after} the derivation of
the field equation by means of the variational principle with the
Lagrangian \eqref{U1Lagrangian}.

\subsection{A quartet of massive one-form fields}

If we used massive fields, then we would have no difficulties with
lack of gauge invariance, because the mass term is not gauge invariant
anyway. Since we want to study gravity \`a la Kaniel-Itin, we start
with a quartet of 1-form fields $k^I$, where $I$ is an {internal}
index with $I=\check{0},\check{1},\check{2},\check{3}$. We again
derive a wave type equation as in the last subsection, but we now add
a massive term for each of the four fields:
\begin{equation}\label{k-lagrangian}
  L_k=\frac{1}{2}\left(dk^I\wedge{}^\star dk^I +\;d^\dagger
    k^I\wedge{}^\star d^\dagger k^I-m_{(I)}\,k^I\wedge{}^\star
    k^I\right)+L_{\text{mat}} \,.\end{equation} We vary with respect to
$k^I$ and find as the Euler-Lagrange equation:
\begin{equation}\label{k-fieldeq}
  \left(\square +m_{(I)}\right){}^\star k^I= \frac{\delta
    L_{\text{mat}}}{\delta k^I} \,.\end{equation} One could also think
of an additional Higgs-type (or `cosmological') term.  Then we would
have
\begin{equation}\label{k-lagrangian'}\begin{split}
  L_{k'}=\frac{1}{2}\left(dk^I\wedge{}^\star dk^I\right.&\left. +\;d^\dagger
    k^I\wedge{}^\star d^\dagger
    k^I-m_{(I)}\,k^I\wedge{}^\star
    k^I\right)\\&-\frac{\lambda}{4!}\,{\epsilon}_{IJKL}\,k^I\wedge
  k^J\wedge k^K\wedge k^L+L_{\text{mat}} \,,\end{split}\end{equation}
and, as field equation,
\begin{equation}\label{k-fieldeq'}
  \left(\square +m_{(I)}\right){}^\star
  k^I+\frac{\lambda}{3!}\,\epsilon_{IJKL}\, k^J\wedge k^K\wedge k^L
  =\frac{\delta L_{\text{mat}}}{\delta k^I} \,.\end{equation} 

This is as near as we can approach the field equation \eqref{WAVE}.
Since currents are 3-forms, we take the Hodge dual of \eqref{WAVE} and
remember $^\star\,\square=\square\,^\star$. Furthermore we put a
source term on its right hand side. In gravitational theory, this can
be only the matter current $\Sigma_\alpha$, representing the
energy-momentum flux of matter. Then the completed Kaniel-Itin field
equation reads:
\begin{equation}\label{KI-fieldeq}\left[\square +\lambda(x)\right]{}^\star
  \vartheta_\alpha=\frac{\delta L_{\text{mat}}}
  {\delta\vartheta^\alpha} =:\Sigma_\alpha\,.\end{equation} Since
$\lambda(x)$ is a \emph{function}, it cannot be identified with some
constant mass $m_{(I)}$. Also an interpretation of $\lambda(x)$ as a
cosmological constant is obviously meaningless. Therefore the
equations \eqref{k-fieldeq'} and \eqref{KI-fieldeq} have to be
carefully distinguished. In future, we will refer to \eqref{KI-fieldeq}
as the (completed) Kaniel-Itin field equation. Eq.\eqref{KI-fieldeq}
represents the heart of their theory.

\subsection{Relation to teleparallel theories}\label{sec:KI+tele}

Kaniel-Itin proposed in \cite[Eq.(8)]{kani97} the following Lagrangian
for the derivation of the vacuum version of \eqref{KI-fieldeq}:
\begin{equation}
  \stackrel{\text{orig}}{V_{\rm{KI}}} \,=\, \frac{1}{2}\left[
    d\vartheta^\alpha \wedge \ensuremath{{}^\star} d\vartheta_\alpha -
    d^\dagger \vartheta^\alpha\wedge\ensuremath{{}^\star} d^\dagger
    \vartheta_\alpha+\lambda(x)(\vartheta^\alpha\wedge{}^\star
    \vartheta_\alpha-4\eta)\right]\,.
\end{equation}
We used our notation here. We want to correct this Lagrangian in two
respects: (i) If we compare \eqref{k-lagrangian'} with
\eqref{k-fieldeq'}, it is clear that we should \emph{add} the first
two terms, instead of subtracting them. (ii) The Lagrange multiplier
term is a hoax since the expression multiplying $\lambda(x)$
\emph{vanishes} identically: We have quite generally
$\vartheta^\alpha\wedge\eta_\beta=\delta^\alpha_\beta\eta$, with
$\eta_\beta:={}^\star\vartheta_\beta$, and the trace of this equation
proves our contention. Taking care of both objections, we will call
\begin{equation} V_{\rm{KI}} \,=\,
  \frac{1}{2}\left( d\vartheta^\alpha \wedge \ensuremath{{}^\star}
    d\vartheta_\alpha + d^\dagger
    \vartheta^\alpha\wedge\ensuremath{{}^\star} d^\dagger
    \vartheta_\alpha\right)\,\label{KI-lagrangian} \end{equation} the
(corrected) Kaniel-Itin Lagrangian. If we recall that we were able to
derive \eqref{k-fieldeq'} from \eqref{k-lagrangian'} only because the
$k^I$ was an internal field, the variation of which commutes with the
star, then it becomes clear that Kaniel-Itin field equation
{\em (\ref{KI-fieldeq})}, for $\Sigma_\alpha=0$, \emph{is not the Euler-Lagrange
  equation of} {\em (\ref{KI-lagrangian})}.

In the Kaniel-Itin model, the Hodge star {no longer commutes} with the
variation $\delta\vartheta^\alpha$ since their gravitational potential
$\vartheta^\alpha$ is inseparably connected to the spacetime manifold.
A postulate of Kaniel-Itin to the opposite, see \cite[statement
between Eqs.(7) and (8)]{kani97}, is without foundation, at least from
a geometrical point of view. Consequently, we will give up this
postulate. The Kaniel-Itin Lagrangian belongs to the so-called
teleparallelism models of gravity. We will come back to this in Sec.\ 
\ref{sec:tele}.

The addition of the adjoint piece in \eqref{KI-lagrangian} does {not}
break the translational invariance since
\(d\ensuremath{{}^\star}\vartheta^\alpha\), and thus
\(d^{\dagger}\vartheta^\alpha\), can be rewritten in terms of
\(d\vartheta^\alpha\) (shown here for an orthonormal coframe):
\begin{equation} d\,\ensuremath{{}^\star} \vartheta^\alpha =
  d\eta^\alpha = d\vartheta_\beta \wedge \eta^{\alpha\beta}
\label{eq:d-eta}\;. \end{equation} 
An analogous procedure is \emph{not} available in the Maxwellian case
for $d{}^\star A$, that is, $d{}^\star A$ cannot be expressed in terms
of $dA$. Therefore $L_{\rm Max}^\dagger$ is not gauge invariant, see
\eqref{gaugevariance}, and has to be rejected as a decent Lagrangian.

\section{Variation of the Hodge dual of a form}\label{sec:hodge}
\subsection{The master formula}
In order to install the \emph{variation} $\delta$ as a derivation, we
demand that it fulfills an even Leibniz rule,\label{sec:rules}
\begin{equation}
\delta(\omega_1 \wedge \omega_2) = \delta\omega_1 \wedge \omega_2 +
\omega_1 \wedge \delta\omega_2 \;,\label{eq:var-leibniz}
\end{equation}
where \(\omega_1\) and \(\omega_2\) are arbitrary exterior
differential forms.  The Leibniz rule is even, because the variation
does not change the degree of the form.
In contrast to this, the interior product \(v\rfloor\) (here $v$ is a
vector) and the exterior derivative \(d\) decrease or increase,
respectively, the degree of the form by one and fulfill an \emph{odd}
Leibniz rule.

Furthermore, we need a relation between the variation and the exterior
derivative. According to the definition of the variation, they simply
commute:
\begin{equation} [d,\delta] = 0\;. \label{eq:var-comm} \end{equation}
Let us now turn to the Hodge star operator, see \cite{hehl95}. It maps
a $p$--form
$\psi=\frac{1}{p!}\,\psi_{\alpha_1\cdots\alpha_p}\vartheta^{\alpha_1}
\wedge\cdots\wedge\vartheta^{\alpha_p}$ into an $(n-p)$--form
$^\star\psi\,$; here $n$ is the dimension of the manifold, i.e., in our
case $n=4$. In terms of components we have
\begin{multline}\label{Hodgestar}
    ^\star\psi:=  \frac{1}{(n-p)!\, p!}\, \sqrt{|\det g_{\mu\nu}|}\,
    g^{\alpha_1\gamma_1}\cdots g^{\alpha_p\gamma_p}\,\times\\
     \epsilon_{\alpha_1\cdots\alpha_p\beta_1\cdots\beta_{n-p}}\,
    \psi_{\gamma_1\cdots\gamma_p}\,
    \vartheta^{\beta_1}\!\wedge\cdots\wedge\vartheta^{\beta_{n-p}}\,,
\end{multline}
where $\epsilon$ is the Levi-Civita symbol. Besides the
$\vartheta$-basis $\Bigl\{{ 1},\vartheta^{\alpha_1},
\vartheta^{\alpha_1}\wedge\vartheta^{\alpha_2},\dots,
\vartheta^{\alpha_1}\wedge\vartheta^{\alpha_2}\wedge\dots
\wedge\vartheta^{\alpha_n}\Bigr\}$, having the Hodge star at our
disposal, we may define the so-called $\eta$-basis:
\begin{equation}\label{etabasis}\begin{split}&\Bigl\{\eta,\eta^{\alpha_1}, 
  \eta^{\alpha_1\alpha_2},\dots,\eta^{\alpha_1\alpha_2\cdots
    \alpha_n}\Bigr\}:=\\ &  \qquad\Bigl\{{}^\star{ 1},{}^\star
  \vartheta^{\alpha_1},
  {}^\star(\vartheta^{\alpha_1}\wedge\vartheta^{\alpha_2}),\dots,
  {}^\star(\vartheta^{\alpha_1}\wedge\vartheta^{\alpha_2}\wedge\dots
  \wedge\vartheta^{\alpha_n})\Bigr\}\,.\end{split}
\end{equation}

Now we can derive the desired expression for
\(\delta\ensuremath{{}^\star}\phi\), for an arbitrary \(p\)-form
\(\phi\). We can saturate the $(n-p)$-form $^\star\phi$ with coframes
$\vartheta^\beta$ such as to arrive at the $n$-form
\newcommand{\ptetr}{\ensuremath{\vartheta^{\beta_1}\wedge \dots
\wedge \vartheta^{\beta_p}}}%
\begin{equation}\label{vorher}
  \ptetr \wedge \ensuremath{{}^\star} \phi
  \stackrel{\eqref{eq:hodge-wedge}}{=} \phi \wedge
  \ensuremath{{}^\star} \left(\ptetr\right)
  \stackrel{\eqref{eq:eta-def}}{=} \phi \wedge
  \eta^{\beta_1\dots\beta_p} \;.\end{equation} We vary \eqref{vorher}.
Then the even Leibniz rule \eqref{eq:var-leibniz} for the variation
leads to
\begin{multline}\label{varphi}
    \delta \left( \ptetr \right)
\wedge \ensuremath{{}^\star} \phi +
    \ptetr \wedge \delta \ensuremath{{}^\star} \phi = \\ 
\delta \phi
    \wedge \eta^{\beta_1\dots\beta_p} + \phi \wedge \delta
    \eta^{\beta_1\dots\beta_p} \;.
\end{multline} Thus,
apparently, we know how to vary the Hodge star, provided we know how
to vary the $\eta$-basis. The variation of the $(n-p)$-form
$\eta^{\alpha_1\cdots\alpha_p}$ is computed in the appendix in Sec.\ 
\ref{sec:2.2}. It turns out to be
\begin{equation}
\label{eq:var-eta} \delta \eta^{\beta_1\dots\beta_p} =
\delta\vartheta^\mu \wedge \left( e_\mu \rfloor
  \eta^{\beta_1\dots\beta_p} \right) + \delta
g_{\kappa\lambda}\left( \vartheta^{(\kappa |} \wedge
  \eta^{\beta_1 \dots \beta_p | \lambda)} - \frac{1}{2} \,
   g^{\kappa\lambda} \eta^{\beta_1\dots\beta_p} \right) \,.
\end{equation}
Incidentally, for the special choice of an orthonormal tetrad, we have
\(\delta g_{\alpha\beta}=0\), and the last two terms vanish; in
particular, we then have \(\delta \eta^{\beta_1\dots\beta_n}=0\). But we
will not introduce this specialization at the present stage.

If we resolve (\ref{varphi}) with respect to $\delta
\ensuremath{{}^\star} \phi$, then, after some intermediary algebra,
see Sec.\ \ref{deduc33} of the appendix, we find for arbitrary
\(p\)-forms \(\phi\) the master formula:
\begin{equation}
  \boxed{\begin{split} \left(\delta \ensuremath{{}^\star}
        -\ensuremath{{}^\star} \delta\right) \phi & = \;\delta
      \vartheta^\alpha \wedge \left(e_\alpha \rfloor
        \ensuremath{{}^\star} \phi \right) - \ensuremath{{}^\star}
      \big[ \delta \vartheta^\alpha \wedge \left( e_\alpha \rfloor
        \phi \right)\big] \label{eq:var-hodge} \\ & + \delta
      g_{\alpha\beta}\Big[\vartheta^{(\alpha}\! \wedge\!
      ( e^{\beta)}
 \rfloor \ensuremath{{}^\star} \phi ) -
      \frac{1}{2}\,g^{\alpha\beta}\;
      \ensuremath{{}^\star} \phi\Big] \;.
\end{split}}
\end{equation}
Again, for {orthonormal} (co)frames, the terms in the second line
vanish since \(\delta g_{\alpha\beta}=0\).

\subsection{Constrained variations \`a la Yang-Mills}\label{sec:constraint}
\textbf{Proposition}: \emph{The condition
\begin{equation}
\delta\ensuremath{{}^\star}\phi=\ensuremath{{}^\star}\delta\phi
\label{eq:comm}\end{equation} 
for an arbitrary \(p\)-form \(\phi\) is equivalent to
the following relation between the variation of the metric and the coframe:
\begin{equation}\label{constraint}
  \delta g_{\alpha\beta}=-2g_{\gamma (\alpha}\,e_{\beta
    )}\rfloor\delta\vartheta^\gamma =-2\omega_{(\alpha\beta)}\,,\qquad
  \text{with}\qquad
  \delta\vartheta^\gamma=\omega_\delta{}^\gamma\vartheta^\delta
  \,.\end{equation} Therefore, for an orthonormal coframe, the allowed
variations are of the Lorentz type, i.e.,
$\omega_{(\alpha\beta)}\equiv 0.$}

To prove\footnote{It was Yuri Obukhov who suggested essential parts of
  this proof to us.} this equivalence we first assume that
\eqref{eq:comm} is valid. We apply this constrained variation to the
volume $n$-form \(\eta:={}^\star 1\), defined in \eqref{etabasis},
\begin{equation} \delta\eta = \delta(\ensuremath{{}^\star} 1) = 
  \ensuremath{{}^\star}(\delta 1) \equiv 0\;,\label{P1}
\end{equation}
since the constant 1 is not varied. 
In turn, for the identity
\begin{equation} \vartheta^\alpha\wedge\eta_\beta = \delta^\alpha_\beta\,\eta
\label{P2}\end{equation} 
we find:
\begin{equation} \delta\vartheta^\alpha\wedge\eta_\beta +
\vartheta^\alpha\wedge\delta\eta_\beta = 0\;.\label{P3}
\end{equation}
The commutation rule \eqref{eq:comm} applied to \(\vartheta_\beta\)
yields
\begin{equation} \delta\eta_\beta = \delta(\ensuremath{{}^\star} 
  \vartheta_\beta)
  =\ensuremath{{}^\star}(\delta\vartheta_\beta)\;.\end{equation} Thus,
by exterior multiplication with $\vartheta^\alpha$ we arrive at
\begin{equation}
  \vartheta^\alpha\wedge\delta\eta_\beta =
  \vartheta^\alpha\wedge\ensuremath{{}^\star} (\delta\vartheta_\beta)=
  \delta\vartheta_\beta\wedge\ensuremath{{}^\star}\vartheta^\alpha=
  \delta(g_{\beta\gamma}\vartheta^\gamma)\wedge\eta^\alpha\;.\end{equation}
On substitution into \eqref{P3}, we find
\begin{equation} \delta\vartheta^\alpha\wedge\eta_\beta +g_{\beta\gamma}\,
  \delta\vartheta^\gamma\wedge\eta^\alpha+\delta
  g_{\beta\gamma}\,\vartheta^\gamma\wedge \eta^\alpha =0 \label{P4}
\end{equation}or, since $\eta_\beta=e_\beta\rfloor\eta$, 
\begin{equation}\label{strain}
  \delta g_{\alpha\beta}=-2g_{\gamma(\alpha}\,e_{\beta)}\rfloor\delta
  \vartheta^\gamma\,.\end{equation} The 1-form
$\delta\vartheta^\alpha$ can be expanded with respect to the coframe:
\begin{equation} \label{eq:var-exp} \delta \vartheta^\gamma =
  \omega_\delta{}^\gamma \vartheta^\delta \,.\end{equation} We insert
\eqref{eq:var-exp} into \eqref{strain}. Then we find
\begin{equation} \label{previous}
  \delta g_{\alpha\beta}=-2 \omega_{(\alpha\beta)} \,.
\end{equation}

To investigate the reverse part of the proposition, we apply the
general rule \eqref{eq:var-hodge} for the variations of Hodge dual
forms and use \eqref{eq:var-exp} and \eqref{previous}:
\begin{equation}
  \begin{split} \left(\delta \ensuremath{{}^\star}
      -\ensuremath{{}^\star} \delta\right) \phi & =
    \;\omega^{\beta\alpha}\Big[\vartheta_\beta \wedge \left(e_\alpha
      \rfloor \ensuremath{{}^\star} \phi \right)-
    \ensuremath{{}^\star} \big[\vartheta_\beta \wedge \left( e_\alpha
      \rfloor \phi \right)\big]\Big] \label{eq:var-hodge-sub} \\ 
    &\quad -2 \omega^{(\beta\alpha)}\,\vartheta_{\alpha}\! \wedge\!  (
    e_{\beta} \rfloor \ensuremath{{}^\star} \phi
    )+\omega_\gamma{}^\gamma\; \ensuremath{{}^\star} \phi\\&=
    \;\omega^{\beta\alpha}\Big[-\vartheta_\alpha \wedge \left(e_\beta
      \rfloor \ensuremath{{}^\star} \phi \right)-
    \ensuremath{{}^\star} \big[\vartheta_\beta \wedge \left( e_\alpha
      \rfloor \phi \right)\big]+g_{\alpha\beta}\,\ensuremath{{}^\star}
    \phi\Big] \\&=
    \;\omega^{\beta\alpha}\Big[e_\beta\rfloor(\vartheta_\alpha \wedge
    \ensuremath{{}^\star} \phi )- \ensuremath{{}^\star}
    \big[\vartheta_\beta \wedge \left( e_\alpha \rfloor \phi
    \right)\big]\Big]\\ & \stackrel{\text{\eqref{eq:hodge-coframe1},
        \eqref{eq:hodge-coframe3}}}{=}
    \omega^{\beta\alpha}\,e_\beta\rfloor{}^\star(e_\alpha\rfloor\phi)\Big[
    \frac{1}{(-1)^{p-1}}-(-1)^{p-1}\Big]= 0\;.
\end{split}
\end{equation}
Thus the proposition is proved.

\section{Teleparallelism theories of gravity}\label{sec:tele}

A Minkowski space is invariant under rigid translations. In order to
make a manifold locally translation invariant, one can introduce a
gauge potential by means of a dynamical coframe \(\vartheta^\alpha\),
see \cite{gron97,gron96,egg1,eggannals,haynew,Thirring,Wallner}. Such
a spacetime carries a torsion, but no curvature: It is a so-called
\emph{Weitzenb\"ock spacetime}. Then, by picking a suitable frame, the
connection \(\Gamma_\alpha{}^\beta\) can always globally be
transformed to zero.

Therefore, in a teleparallel theory, the coframe \(\vartheta^\alpha\)
is the basic gravitational field variable. Furthermore, let a metric
\(g\) be given of Minkowskian signature, i.~e.~of index 3 (number of
negative eigenvalues of the metric). For the rest of this paper we
choose the coframe to be \emph{orthonormal}, \(g := o_{\alpha\beta}\,
\vartheta^\alpha \otimes \vartheta^\beta\), with \(o_{\alpha\beta} =
\text{diag}(+1, -1, -1, -1)\), and raise and lower the frame indices
by means of \(o_{\alpha\beta}\).

\subsection{The Rumpf Lagrangians}
According to Rumpf \cite{rumpf78}, a general quadratic Lagrangian for
the coframe \(\vartheta^\alpha\) can be expanded in terms of the
\emph{gauge-invariant} translational Lagrangians ($\ell=$ Planck
length, $\Lambda=$ cosmological constant $=\rho_0/2$):
\begin{equation} 
  V = \frac{1}{2\ell^2}\sum_{K=0}^4 \rho_{K} \; {}^{[K]}V\;,
  \label{eq:decomp-rumpf}
\end{equation}
with 
\begin{subequations}
\label{eq:rumpf}
\begin{align}
  \,{}^{[0]}V &= \frac{1}{4} \, \vartheta^\alpha \wedge
  \ensuremath{{}^\star} \vartheta_\alpha = \eta \;,\\ \,{}^{[1]}V &=
  d\vartheta^\alpha \wedge \ensuremath{{}^\star} d\vartheta_\alpha
  \qquad \text{(pure Yang-Mills type)}\;, \\\, {}^{[2]}V &=
  \Big(d\vartheta_\alpha \wedge \vartheta^\alpha \Big) \wedge
  \ensuremath{{}^\star} \left( d\vartheta_\beta \wedge \vartheta^\beta
  \right) \;, \\\, {}^{[3]}V & = \left(d\vartheta^\alpha \wedge
    \vartheta^\beta \right) \wedge \ensuremath{{}^\star} \left(
    d\vartheta_\alpha \wedge \vartheta_\beta \right) =
  d\vartheta^\alpha \wedge \vartheta^\beta \wedge \left( e_\beta
    \rfloor \ensuremath{{}^\star} d\vartheta_\alpha \right) = 2 \,
  {}^{[1]}V \;, \\\, {}^{[4]}V &= \left(d\vartheta_\alpha \wedge
    \vartheta^\beta \right) \wedge \ensuremath{{}^\star} \Big(
  d\vartheta_\beta \wedge \vartheta^\alpha \Big) \;.
\end{align}
\end{subequations}
Since \({}^{[3]}V =2 \,{}^{[1]}V\), we can always put \(\rho_{3}=0\).  

\subsection{The irreducible Lagrangians}
Alternatively, the field strength \(d\vartheta^\alpha\) can be
decomposed into three pieces which transform irreducibly under the
Lorentz group:
\begin{equation}
d\vartheta^\alpha\,=\,{}^{(1)}d\vartheta^\alpha +
{}^{(2)}d\vartheta^\alpha + {}^{(3)}d\vartheta^\alpha
\;. \label{eq:irred-decomp} 
\end{equation}
Here we defined (in parentheses we are mentioning the
corresponding names of our computer algebra programs):
\begin{subequations}
\label{eq:irred}
\begin{align}
  {}^{(1)}d\vartheta^\alpha &:= d\vartheta^\alpha -
  {}^{(2)}d\vartheta^\alpha-{}^{(3)}d\vartheta^\alpha &&
  \texttt{(tentor),}\\ {}^{(2)}d\vartheta^\alpha &:= {1\over
    3}\vartheta^\alpha\wedge(e_\beta\rfloor d\vartheta^\beta) &&
  \texttt{(trator),}\\ {}^{(3)}d\vartheta^\alpha &:= {1\over 3}
  e_\alpha \rfloor \left(\vartheta^\beta\wedge d \vartheta_\beta\right
  ) && \texttt{(axitor).}
\end{align}
\end{subequations}
In terms of the numbers of components involved, we have the
decomposition \(24 = 16 \oplus 4 \oplus 4\). 

Then we can write
\begin{equation} V =\frac{1}{2\ell^2}\left[a_0\, 
    \eta+\sum_{I=1}^3 a_I \left( d\vartheta^\alpha \wedge
      \ensuremath{{}^\star} \, {}^{(I)}d\vartheta_\alpha \right)
  \right]\;. \label{eq:decomp-irred}
\end{equation}

We substitute \eqref{eq:irred-decomp} and \eqref{eq:irred} into
\eqref{eq:rumpf}. Then a comparison between \eqref{eq:decomp-rumpf}
and \eqref{eq:decomp-irred} yields
\begin{align}
\label{eq:rho-a} \rho_{1} & = \frac{1}{3} \, \left(a_2 + 2 a_1 \right)  \;,& 
\rho_{2} & = \frac{1}{3} \, \left(a_3 - a_1 \right) \;,&
\rho_{4} & = \frac{1}{3} \, \left(a_1 - a_2 \right) \;,\\ 
\label{eq:a-rho} a_1 & = \rho_{1} + \rho_{4} \;,&
a_2 & = \rho_{1} - 2 \rho_{4} \;, &
a_3 & = \rho_{1} + 3 \rho_{2} + \rho_{4} \;,
\end{align}
and, additionally \(a_0=\rho_0=2\Lambda\). 
These relations were checked by means of a computer algebra program, see 
Sec.\ \ref{sec:ca} of the appendix.

\subsection{Field equation}
The field equation of a general translation invariant Lagrangian reads
\cite{gron97}
\begin{equation} dH_\alpha - E_\alpha = \Sigma_\alpha
\;,\label{eq:field-orig}\end{equation}
with \begin{equation} H_\alpha : = -\frac{\partial V}{\partial
d\vartheta^\alpha}\;,  \quad E_\alpha 
:= \frac{\partial V}{\partial \vartheta^\alpha}\;, \quad 
\text{and} \qquad \Sigma_\alpha := \frac{\delta L_{\text{mat}}}{\delta
\vartheta^\alpha} \label{eq:definis}\;. \end{equation}
In \eqref{eq:definis}, the partial derivatives are implicitly defined by
means of the variation of the Lagrangian:
\begin{equation} \delta V = \delta \vartheta^\alpha \wedge
\frac{\partial V}{\partial 
\vartheta^\alpha} + \delta\, d\vartheta^\alpha \wedge \frac{\partial V}{\partial
d\vartheta^\alpha} \;. \end{equation}

If we use the abbreviations \eqref{eq:definis}, we don't need our
master formula for the computation of the field equation.
Alternatively, one can take the Lagrangian \eqref{eq:decomp-rumpf}
together with \eqref{eq:rumpf} and vary the resulting expression by
using \eqref{eq:var-hodge} with $\delta g_{\alpha\beta}=0$. This
yields the explicit form of the field equation \eqref{eq:field-orig},
cf.\ Kopczy\'nski \cite{Wojtek}\footnote{Kopczy\'nski denoted the
  Rumpf Lagrangians by $K$. We have the following translation rules:
  $K^1={}^{[4]}V$, $K^2={}^{[2]}V$, $K^3={}^{[1]}V$. In
  \eqref{genLag}, the second derivatives of the coframe are exactly
  the same (for $\ell^2=1$) as those in the corresponding three
  equations of Kopczy\'nski \cite[top of p.\ 503]{Wojtek}.}:
\begin{equation}\label{genLag}
\begin{split}
  - 2\,&\ell^2\Sigma_\alpha = 2\rho_1 \; d\,{}^\star d\vartheta_\alpha
  - 2\rho_2\; \vartheta_\alpha \wedge d \;{}^\star\left(
    d\vartheta^\beta\wedge \vartheta_\beta\right) - 2\rho_4\;
  \vartheta_\beta \wedge d \; {}^\star \left( \vartheta_\alpha \wedge
    d \vartheta^\beta \right) \\ & + \rho_1 \Big[ e_\alpha \rfloor
  \left( d\vartheta^\beta \wedge {}^\star d\vartheta_\beta \right) - 2
  \left( e_\alpha \rfloor d\vartheta^\beta \right) \wedge {}^\star
  d\vartheta_\beta\Big] \\ & + \rho_2\;\Big[2 d\vartheta_\alpha \wedge
  {}^\star \left( d\vartheta^\beta \wedge \vartheta_\beta \right) \\&
 \quad + e_\alpha \rfloor \left( d\vartheta^\gamma \wedge
    \vartheta_\gamma \wedge {}^\star \left( d\vartheta^\beta \wedge
      \vartheta_\beta\right) \right) - 2\left( e_\alpha \rfloor
    d\vartheta^\beta\right) \wedge \vartheta_\beta \wedge {}^\star
  \left( d\vartheta^\gamma \wedge \vartheta_\gamma\right)\Big] \\ & +
  \rho_4\Big[2 d\vartheta_\beta \wedge {}^\star \left( \vartheta_\alpha
    \wedge d \vartheta^\beta\right)\\&\quad + e_\alpha \rfloor
  \left( \vartheta_\gamma \wedge d\vartheta^\beta \wedge {}^\star
    \left( d\vartheta^\gamma \wedge \vartheta_\beta \right)\right) - 2
  \left( e_\alpha \rfloor d\vartheta^\beta\right) \wedge
  \vartheta_\gamma \wedge {}^\star \left( d\vartheta^\gamma \wedge
    \vartheta_\beta \right)\Big] \,.
\end{split}
\end{equation}
In the first line of this equation we displayed the leading terms
containing second derivatives of the coframe. In the remaining terms
there enter only first derivatives.

\subsection{Decomposition of the Kaniel-Itin Lagrangian}
In order to better recognize the structure of the KI-Lagrangian, one
can decompose it in its irreducible pieces as well as into the Rumpf
Lagrangians.\label{sec:KI-decomp}

The first term of the KI-Lagrangian \eqref{KI-lagrangian} is exactly
the Yang-Mills type Lagrangian \({}^{[1]}V\), cf.\ with
\eqref{eq:rumpf}.  The term \(d^{\dagger} \vartheta^\alpha \wedge
\ensuremath{{}^\star} d^{\dagger} \vartheta_\alpha =
-d\ensuremath{{}^\star} \vartheta^\alpha \wedge \ensuremath{{}^\star}
d\ensuremath{{}^\star} \vartheta_\alpha\) is left for scrutiny. Using
formula \eqref{eq:d-eta} and other rules, see, e.g., \cite{hehl95} and
Sec.\ \ref{formulae} for details, we find
\begin{equation}
\begin{split}
  -d\,\ensuremath{{}^\star} \vartheta^\alpha \wedge
  \ensuremath{{}^\star} d\, \ensuremath{{}^\star} \vartheta_\alpha & =
  -d\eta^\alpha \wedge \ensuremath{{}^\star} d\eta_\alpha =
  -d\vartheta_\beta \wedge \eta^{\alpha\beta} \wedge
  \ensuremath{{}^\star} \left( d\vartheta^\gamma \wedge
    \eta_{\alpha\gamma} \right) \\ & = -\vartheta^\alpha \wedge
  \vartheta^\beta \wedge \ensuremath{{}^\star} d\vartheta_\beta \wedge
  \ensuremath{{}^\star} \left( \vartheta_\alpha \wedge
    \vartheta_\gamma \wedge \ensuremath{{}^\star} d\vartheta^\gamma
  \right) \\ & = \vartheta^\alpha \wedge \vartheta^\beta \wedge
  \ensuremath{{}^\star} d\vartheta_\beta \wedge \ensuremath{{}^\star}
  \left[ \vartheta_\alpha \wedge \ensuremath{{}^\star} \left(e_\gamma
      \rfloor d\vartheta^\gamma \right)\right] \\ & = \vartheta^\beta
  \wedge \ensuremath{{}^\star} d\vartheta_\beta \wedge
  \vartheta^\alpha \wedge \left[ e_\alpha \rfloor \left( e_\gamma
      \rfloor d\vartheta^\gamma \right)\right] \\ & =
  \ensuremath{{}^\star} d\vartheta_\beta \wedge \vartheta^\beta \wedge
  \left( e_\gamma \rfloor d\vartheta^\gamma\right) =
  \ensuremath{{}^\star} \left[ \vartheta^\beta \wedge \left( e_\gamma
      \rfloor d\vartheta^\gamma\right) \right] \wedge
  d\vartheta_\beta\\ & = - d\vartheta_\beta \wedge
  \ensuremath{{}^\star} \left[ e_\gamma \rfloor \left( \vartheta^\beta
      \wedge d\vartheta^\gamma\right)\right] + d\vartheta_\gamma
  \wedge \ensuremath{{}^\star} d\vartheta^\gamma \\ &= -
  d\vartheta_\alpha \wedge \vartheta_\beta \wedge
  \ensuremath{{}^\star} \left( \vartheta^\alpha \wedge
    d\vartheta^\beta \right) + d\vartheta^\alpha \wedge
  \ensuremath{{}^\star} d\vartheta_\alpha = {}^{[1]}V - {}^{[4]}V\;.
\end{split}
\end{equation}
Accordingly, the KI-Lagrangian \eqref{KI-lagrangian} can be rewritten
as
\begin{equation}
\label{eq:base-rumpf} 
V_{\text{KI}} = \frac{1}{2\ell^2}\Big[2\, d\vartheta^\alpha \wedge
\ensuremath{{}^\star} d\vartheta_\alpha - \left(d\vartheta_\alpha 
\wedge \vartheta^\beta \right) \wedge \ensuremath{{}^\star}
\Big( d\vartheta_\beta \wedge \vartheta^\alpha \Big)\Big]\;, 
\end{equation}
and we can read off the \(\rho_K\) coefficients as follows:
\begin{equation}\label{KIcoeff}
\rho_{1} = 1+1 = 2\;, \quad
\rho_{2} = 0 \;, \quad 
\rho_{4} = - 1 \;.
\end{equation}
By \emph{subtracting} the adjoint term we would have
\(\rho_1=\rho_2=0\) and \(\rho_4=1\), i.e., we would get the von der
Heyde Lagrangian \cite{heyde76}.

The coefficients of the decomposition into irreducible pieces, by 
using \eqref{eq:a-rho}, turn out to be
\begin{equation}
  a_1 = 1 \;, \quad a_2=4 \;, \quad a_3 = 1\label{eq:coeff-ki}
\end{equation}
(in the von der Heyde case, we have \(a_1=1\), \(a_2= -2\), \(a_3=1\)). 
Accordingly, the KI-Lagrangian can be rewritten in the form
\begin{equation}
V_{\text{KI}}= -\frac{1}{2}\, d\vartheta^\alpha\wedge H_\alpha \;,
\label{eq:base-irred} \end{equation}
with the translational ``excitation''
\begin{equation} 
  H_\alpha = -\frac{1}{\ell^2} \,\ensuremath{{}^\star}
  \!\!\left(a_1\,{}^{(1)}d\vartheta_\alpha+
    a_2\,{}^{(2)}d\vartheta_\alpha+a_3\,{}^{(3)}d\vartheta_\alpha\right)
\end{equation}
and the coefficients \eqref{eq:coeff-ki}.

The Lagrangian \(V_{\text{KI}}\) is \emph{not}
locally Lorentz invariant.  Rather, a locally Lorentz invariant theory results
from the following choice of the parameters:
\begin{equation}
a_1\, =\, 1\;, \quad a_2\,=\, -2 \;,  \quad a_3 = -{1\over 2}\;.
\label{eq:coeff-gr} 
\end{equation}
This represents the teleparallel
\emph{equivalent} of Einstein's general relativity.

\subsection{Viable Lagrangians}
The form of a general quadratic Lagrangian was displayed in Eqs.\ 
(\ref{eq:decomp-irred}) and (\ref{eq:decomp-rumpf}).  Various choices
of parameters $a_I$ or $\rho_K$ correspond to various teleparallel
theories of gravity. We call a specific Lagrangian \emph{viable} if it
leads to a theory which fulfills the following conditions: (i) It has
the correct Newtonian approximation. (ii) It agrees with the first
post-Newtonian approximation of general relativity. (iii) It has the
Schwarzschild metric as exact solution in the case of spherical
symmetry. 

The question which parameters yield a viable Lagrangian has already
been discussed in the literature, see for example \cite{Nitsch,
  straumann2, Wojtek,Thirring}. The result is that we have viable
Lagrangians for $a_1=1\,,\;a_2=-2\,,\;a_3=\text{arbitrary}$ or
$\rho_4=1\,,\;\rho_1=0\,,\;\rho_2=\text{arbitrary}$. The arbitrary
$a_3$ or $\rho_2$ pieces, respectively, represents the axial square
contribution $A\wedge{}^\star A $ of the torsion, with
$A:=\frac{1}{3}\,\vartheta^\beta\wedge d\vartheta_\beta$ and
${}^{(3)}d\vartheta^\alpha=e_\alpha\rfloor A$. Deviations between
viable theories due to different axial pieces only show up in fifth
order of the post-Newtonian approximation \cite{Nitsch}.  Therefore,
on a phenomenological level, all viable teleparallel theories are
indistinguishable. In Table 1 we have listed some quadratic torsion
Lagrangians.

It follows from the end of the last subsection, see also
\eqref{eq:coeff-gr}, that only the teleparallel equivalent of
Einstein's general relativity is both viable and locally Lorentz
invariant. It has not yet been clearly answered if in this context
local Lorentz invariance is obligatory or merely an aesthetic feature.

\begin{table}[htb]
\begin{center}
\begin{tabular}{c|rrrrrr}
  & $\text{GR}_\parallel$ \cite{gron96}& vdH \cite{heyde76} & viable
  & YM &YM$^\dagger$& $\text{KI}$ \cite{kani97}\\ \hline $a_1$ &
  1 & 1 & 1 & 1 & 0 & 1\\ $a_2$ & $-2$ & $-2$ &$-2$  & 1 &3 & 4 \\ 
  $a_3$ & $-\frac{1}{2}$ & 1 &\text{arb.}  & 1& 0& 1\\ \hline $\rho_1$
  & 0 & 0 & 0 & 1 &1& 2\\ $\rho_2$ & $-\frac{1}{2}$ & 0 & \text{arb.} & 0 &0&
  0\\ $\rho_4$ & 1 & 1 & 1 & 0&$-1$&$-1$
\end{tabular}
\end{center}
\caption{\it This table lists the $a_I$ and the $\rho_K$ coefficients for 
  different teleparallel Lagrangians. $\text{GR}_\parallel$, spelled
  out in the first column, represents a viable gravitational model,
  the same is true for the von der Heyde case. Obviously, the Kaniel-Itin 
  Lagrangian, in the   framework of the conventional variational procedure, is
  not viable.   We have $\text{KI}=\text{YM}+\text{YM}^\dagger$ and
  $\text{vdH}= \text{YM}-\text{YM}^\dagger$.}
\end{table}

\subsection{Schweizer-Straumann-Wipf amended}

Schweizer and Straumann \cite{straumann1} and Schweizer, Straumann,
and Wipf \cite{straumann2} investigated the von der Heyde
teleparallelism Lagrangian \cite{heyde76}, see also \cite{Nitsch}.  In
particular they showed that, to first post-Newtonian order, the von
der Heyde theory predicts the same gravitational radiation loss as
general relativity. However, they assumed in some places the incorrect
commutation rule \(\delta\, {}^\star = {}^\star \delta\). Therefore
some equations in these articles must be corrected. We stress that the
corrections do not influence their overall results, though.

We present the corrected formulas of the article by Schweizer,
Straumann, and Wipf \cite{straumann2} in the numbering used there, but
in our notation. The correcting terms are printed in bold.

The wrong variation first shows up in the explicit expressions of
the canonical energy-momentum tensors derived from the teleparallel
version of the Hilbert-Einstein Lagrangian, \(\epsilon_\alpha^E\), and the
difference between the von der Heyde Lagrangian and the teleparallel
version of the Hilbert-Einstein Lagrangian,
\(\Delta \epsilon_\alpha\):
\begin{align}
  \tag*{(2.12),\cite{straumann2}} \epsilon_\alpha^E & = - d\left[
    \vartheta^\beta \wedge \ensuremath{{}^\star} \left(
      d\vartheta_\beta \wedge \vartheta_\alpha \right)\right] -
  d\vartheta^\beta \wedge \ensuremath{{}^\star} \left(
    d\vartheta_\alpha \wedge \vartheta_\beta \right) \\ & \quad \notag
  + \frac{1}{2} d \left\{ \vartheta_\alpha \wedge
    \ensuremath{{}^\star} \left( d\vartheta^\beta\wedge
      \vartheta_\beta \right) \right\} + \frac{1}{2} d\vartheta_\alpha
  \wedge \ensuremath{{}^\star} \left( d\vartheta^\beta\wedge
    \vartheta_\beta \right) \\ & \notag \quad \qquad \qquad
  \boldsymbol{+ \frac{1}{2} e_\alpha \rfloor \left( \vartheta_\gamma
      \wedge d\vartheta^\beta \right) \wedge \ensuremath{{}^\star}
    \left( d\vartheta^\gamma \wedge \vartheta_\beta \right)} \\ &
  \notag \quad \qquad \qquad \boldsymbol{+ \frac{1}{2}
    d\vartheta^\gamma \wedge \vartheta_\beta \wedge e_\alpha \rfloor
    \ensuremath{{}^\star} \left( \vartheta_\gamma \wedge
      d\vartheta^\beta \right)} \\ & \notag \quad \qquad \qquad
  \boldsymbol{- \frac{1}{4} e_\alpha \rfloor \left( d\vartheta^\beta
      \wedge \vartheta_\beta \right) \wedge \ensuremath{{}^\star}
    \left( d\vartheta^\gamma \wedge \vartheta_\gamma \right)} \\ &
  \notag \quad \qquad \qquad \boldsymbol{ - \frac{1}{4}
    d\vartheta^\beta \wedge \vartheta_\beta \wedge e_\alpha \rfloor
    \ensuremath{{}^\star} \left( d\vartheta^\gamma \wedge
      \vartheta_\gamma \right)}\;, \\ \tag*{(2.13),\cite{straumann2}}
  \Delta \epsilon_\alpha & = d \vartheta_\alpha \wedge
  \ensuremath{{}^\star} \left( d\vartheta^\beta\wedge \vartheta_\beta
  \right) - \frac{1}{2} \vartheta_\alpha \wedge d\,
  \ensuremath{{}^\star} \left( d\vartheta^\beta\wedge \vartheta_\beta
  \right) \\ & \notag \quad \qquad \qquad \boldsymbol{- \frac{1}{4}
    e_\alpha \rfloor \left( d\vartheta^\beta \wedge \vartheta_\beta
    \right) \wedge \ensuremath{{}^\star} \left( d\vartheta^\gamma
      \wedge \vartheta_\gamma \right)} \\ & \notag \quad \qquad \qquad
  \boldsymbol{-\frac{1}{4} d\vartheta^\beta \wedge \vartheta_\beta
    \wedge e_\alpha \rfloor \ensuremath{{}^\star} \left(
      d\vartheta^\gamma \wedge \vartheta_\gamma \right)}\;.
\end{align}
Since the additional terms do not influence the antisymmetric part of
\(\Delta \epsilon_\alpha\), we only need to correct the symmetric
part:
\begin{align}
  \tag*{(3.1),\cite{straumann2}} \Delta \epsilon_\alpha^s & = -
  \frac{1}{2}\; \ensuremath{{}^\star} \, \bigg[ \left(
    d\vartheta_\alpha \wedge \vartheta^\beta + d\vartheta^\beta \wedge
    \vartheta_\alpha \right) \wedge \ensuremath{{}^\star} \left(
    d\vartheta^\gamma \wedge \vartheta_\gamma\right) \\ & \notag
  \qquad \qquad \qquad \boldsymbol{ -\frac{1}{2} \vartheta^\beta
    \wedge d\vartheta^\gamma \wedge \vartheta_\gamma \wedge e_\alpha
    \rfloor \ensuremath{{}^\star} \left( d\vartheta^\delta \wedge
      \vartheta_\delta\right)} \\ & \notag \qquad \qquad \qquad \qquad
  \qquad \left. \boldsymbol{ + \frac{1}{2} d\vartheta^\gamma \wedge
      \vartheta_\gamma \wedge \ensuremath{{}^\star} \left(
        d\vartheta^\delta \wedge \vartheta_\delta\right) \,
      \delta^\beta_\alpha} \right] \,\eta_\beta\;. \\\intertext{Hence
    we find the corrected field equation as}
  \tag*{(3.3),\cite{straumann2}} \epsilon_\alpha^E & - \frac{(\lambda
    -1)}{2} \; \ensuremath{{}^\star} \, \bigg[ \left(
    d\vartheta_\alpha \wedge \vartheta^\beta + d\vartheta^\beta \wedge
    \vartheta_\alpha \right) \wedge \ensuremath{{}^\star} \left(
    d\vartheta^\gamma \wedge \vartheta_\gamma\right) \\ & \notag
  \qquad \qquad \qquad \boldsymbol{ -\frac{1}{2} \vartheta^\beta
    \wedge d\vartheta^\gamma \wedge \vartheta_\gamma \wedge e_\alpha
    \rfloor \ensuremath{{}^\star} \left( d\vartheta^\delta \wedge
      \vartheta_\delta\right)} \\ & \notag \qquad \qquad \qquad \qquad
  \qquad \left. \boldsymbol{ + \frac{1}{2} d\vartheta^\gamma \wedge
      \vartheta_\gamma \wedge \ensuremath{{}^\star} \left(
        d\vartheta^\delta \wedge \vartheta_\delta\right) \,
      \delta^\beta_\alpha} \right] \,\eta_\beta = - t^\alpha \;.
\end{align}

{In our units, we have for the $\lambda$-parameter of Schweitzer et
  al.  $\lambda=-2\rho_2=\frac{2}{3}\,(1-a_3)$.} Like the terms
discussed in \cite{straumann2}, the additional terms are at least
quadratic in \(\phi_{\alpha\beta}\) (which is the symmetric part of
\(\Phi_{\alpha\beta}\) in the expansion \(\vartheta^\alpha = dx^\alpha
+ \Phi^{\alpha}{}_\beta\, dx^\beta\)), such that the arguments of
\cite[\S3]{straumann2} remain unchanged.

The formulas \cite[(4.1)]{straumann2} and \cite[(4.6)]{straumann2}
have to be corrected similarly as the last two equations, replacing
\(d\vartheta^\gamma\) by \(T^\gamma\). Since these explicit
expressions are not used in the remainder of \cite[\S4]{straumann2},
the conclusions remain valid therein.

Finally, the quantities \(A^{\mu\nu}\), \(B^{\mu\nu}\), and
\(C^{\mu\nu}\) (formulas \cite[(5.4b)--(5.4d)]{straumann2}) in the
expansion of the quadratic and higher-order terms  
\begin{equation} \tag*{(5.4a),\cite{straumann2}} \epsilon_Q^\alpha =
  \Delta\epsilon_s^\alpha + \left\{ A^{\beta\alpha} + B^{\beta\alpha}
    + C^{\beta\alpha}\right\} \end{equation} would need corrections.
Since these quantities are not used explicitly in \cite{straumann2},
we did not display the exact expressions here.

\section{Kaniel-Itin examined}\label{sec:KI}
\subsection{Lagrangian}
We come back to the Kaniel-Itin Lagrangian \eqref{KI-lagrangian}:
\begin{equation} V_{\rm{KI}} \,=\, 
  \frac{1}{2\ell^2}\left( d\vartheta^\alpha \wedge
    \ensuremath{{}^\star} d\vartheta_\alpha + d^\dagger
    \vartheta^\alpha\wedge\ensuremath{{}^\star} d^\dagger
    \vartheta_\alpha\right)\,.\label{base} \end{equation} Its
Euler-Lagrange equation can be read off from \eqref{genLag} by
substituting the coefficients \eqref{KIcoeff}:
\begin{equation}\label{genLag'}
\begin{split}
  - 2&\ell^2\,\Sigma_\alpha = 2\; \vartheta_\beta \wedge d \; {}^\star
  \left( \vartheta_\alpha \wedge d \vartheta^\beta \right)+4 \;
  d\,{}^\star d\vartheta_\alpha \\ & + 2 e_\alpha \rfloor \left(
    d\vartheta^\beta \wedge {}^\star d\vartheta_\beta \right) - 4
  \left( e_\alpha \rfloor d\vartheta^\beta \right) \wedge {}^\star
  d\vartheta_\beta - 2 d\vartheta_\beta \wedge {}^\star
  \left( \vartheta_\alpha \wedge d \vartheta^\beta\right)\\& -
  e_\alpha \rfloor \left[ \vartheta_\gamma \wedge d\vartheta^\beta
    \wedge {}^\star \left( d\vartheta^\gamma \wedge \vartheta_\beta
    \right)\right] + 2 \left( e_\alpha \rfloor d\vartheta^\beta\right)
  \wedge \vartheta_\gamma \wedge {}^\star \left( d\vartheta^\gamma
    \wedge \vartheta_\beta \right) \;.
\end{split}
\end{equation}
In the first line, we displayed the second derivatives of the
gravitational potential. As we already saw in Table 1, this field
equation is not viable.

If we use the \emph{the constrained variations} in \eqref{base}, then 
we commute $\delta$ and the star $^\star$ and find by simple algebra:
\begin{equation}\label{last}\ell^2\,\delta\left(V_{\rm{KI}} 
    + L_{\rm{mat}}\right)
  =\delta\vartheta^\alpha\wedge\left(-\square\eta_\alpha+
   \ell^2\,\Sigma_\alpha\right) =0\,.\end{equation} Since the
variations are constrained, we have to turn to our proposition and to
use \eqref{constraint}:
$\delta\vartheta^\alpha=\omega_\beta{}^\alpha\, \vartheta^\beta$, with
$\omega_{(\alpha\beta)}=0$. On substitution in \eqref{last}, the field
equation turns out to be proportional to the \emph{antisymmetric} part
of the wave equation,
\begin{equation}
  \vartheta_{[\alpha} \wedge \square\, \eta_{\beta]} =
  \ell^2\,\vartheta_{[\alpha} \wedge \Sigma_{\beta]}
\label{eq:ki-correct}\;, \end{equation}
rather than to the wave equation itself. Accordingly, the constrained
variations also lead to a dead end and we have to turn back our
attention again to the KI-field equation \eqref{KI-fieldeq}.

\subsection{Decomposition of the field equation}

Let us split the full wave equation into its different pieces. For
that purpose, we most conveniently start from the decomposition of the
energy-momentum current\label{sec:44} as a covector-valued 3-form with
16 independent components:
\begin{equation}\label{cut}
  \Sigma_\alpha=\>\overset{\frown}{\Sigma\kern-1.1em\nearrow}_\alpha +
  \frac{1}{2} \vartheta_\alpha \wedge \left(e_\gamma \rfloor
    \Sigma^\gamma \right)+ \frac{1}{4} e_\alpha \rfloor \left(
    \vartheta^\gamma \wedge \Sigma_\gamma \right)\;. \end{equation} Here
$\overset{\frown}{\Sigma\kern-1.1em\nearrow}_\alpha$ is its
\emph{symmetric traceless} part, the second term on the right hand
side its \emph{antisymmetric} and the last term its \emph{trace part},
see \cite[eq.(5.1.15)]{hehl95}.  

We come into better known territory if we decompose $\Sigma_\alpha$
with respect to the $\eta$-basis:
\begin{equation}
  \Sigma_\alpha=T^\beta{}_\alpha\,\eta_\beta \;. \end{equation} The
$T^\beta{}_\alpha$'s are the components of the energy-momentum tensor.
We can `saturate' the 3-form $\Sigma_\beta$ by means of the 1-form
$\vartheta^\alpha$:
\begin{multline} \ensuremath{{}^\star} \left( \vartheta^\alpha \wedge 
    \Sigma_\beta \right) = \ensuremath{{}^\star} \left(
    \vartheta^\alpha \wedge T^\gamma{}_\beta \;\eta_\gamma \right) =
  T^\gamma{}_\beta \; \ensuremath{{}^\star} \left(\vartheta^\alpha
    \wedge \eta_\gamma \right) \\ = T^\alpha{}_\beta \;
  \ensuremath{{}^\star} \eta = T^\alpha{}_\beta
  \;\ensuremath{{}^\star}\ensuremath{{}^\star} 1 = -
  T^\alpha{}_\beta\; \end{multline} or
\begin{equation}
  T_{\alpha\beta}=\ensuremath{{}^\star}\left(\Sigma_\beta
    \wedge\vartheta_\alpha \right)=e_\alpha\rfloor{}^\star\Sigma_\beta
\label{eq:energy-tensor}\;.
\end{equation} 
The analog of \eqref{cut} is, of course, the following splitting of the 
energy-momentum tensor:
\begin{equation}T_{\alpha\beta}=\{T_{(\alpha\beta)}- \frac{1}{4}\,
  T^{\gamma}{}_\gamma\;g_{\alpha\beta}\}
  +T_{[\alpha\beta]}+\frac{1}{4}\,T^{\gamma}{}_\gamma
  \;g_{\alpha\beta}\;.\end{equation}

Coming back to \eqref{KI-fieldeq}, after some algebra, we find the
following decomposition:
\begin{subequations}\label{eq:wave}
\begin{align}
  \square \eta_\alpha - \frac{1}{2} \vartheta_\alpha \wedge
  \left(e_\gamma \rfloor \square\eta^\gamma \right)- \frac{1}{4}
  e_\alpha \rfloor \left( \vartheta^\gamma \wedge \square
    \eta_\gamma\right) & =\>
  \ell^2\,\overset{\frown}{\Sigma\kern-1.1em\nearrow}_\alpha \;,
  \label{eq:ki-symmetric}\\ \frac{1}{2} \vartheta_\alpha \wedge
  \vartheta_\beta \wedge \left( e_\gamma \rfloor \square \eta^\gamma
  \right) & = \ell^2\,\vartheta_{[\alpha} \wedge \Sigma_{\beta]} \;,
  \label{eq:ki-antisymm}\\ \vartheta^\gamma \wedge \square \eta_\gamma
  +4\lambda(x)\,\eta &= \ell^2\,\vartheta^\gamma \wedge \Sigma_\gamma
  \;. \label{eq:ki-trace}
\end{align}
\end{subequations}

We can combine the antisymmetric and the symmetric-tracefree part of
the wave equation in order to get its tracefree part.  The fine
splitting \eqref{eq:wave} simplifies to the tracefree and the trace
part of the field equation (\ref{eq:ki-correct}):
\begin{subequations}
\begin{align} \square \eta_\alpha - \frac{1}{4} e_\alpha \rfloor
  \left( \vartheta^\gamma \wedge \square \eta_\gamma \right)
  &= \ell^2\, \Sigma\kern-1.1em\nearrow_\alpha
  \;,\label{eq:trace-decomp} \\  \vartheta^\gamma \wedge \square
  \eta_\gamma +4\lambda(x)\,\eta
&= \ell^2\,\vartheta^\gamma \wedge \Sigma_\gamma \;.
\label{eq:trace}\end{align}
\end{subequations}

Sometimes it may be useful to rewrite \eqref{eq:trace-decomp} by using
\begin{equation} \begin{split} e_\alpha \rfloor \left( 
      \vartheta^\gamma \wedge \square \eta_\gamma \right) & = -
    e_\alpha \rfloor \ensuremath{{}^\star} \left\{
      \ensuremath{{}^\star} \left( \vartheta^\gamma \wedge \square
        \eta_\gamma \right)\right\} = - \ensuremath{{}^\star} \left(
      \ensuremath{{}^\star} \left( \vartheta^\gamma \wedge \square
        \eta_\gamma \right) \wedge \vartheta_\alpha \right) \\ & = -
    \ensuremath{{}^\star} \left( \vartheta^\gamma \wedge \square
      \eta_\gamma \right) \wedge \ensuremath{{}^\star}
    \vartheta_\alpha = - \ensuremath{{}^\star} \left(\vartheta^\gamma
      \wedge \square \eta_\gamma \right) \eta_\alpha \;,
\end{split} \end{equation}
thereby finding 
\begin{equation}\left[  \square + \frac{1}{4} \ensuremath{{}^\star} 
    \left( \vartheta^\beta \wedge \square \eta_\beta \right) \right]
  \eta_\alpha = \ell^2\,\Sigma\kern-1.1em\nearrow_\alpha\;,
  \label{eq:decomp-second}
\end{equation}or, alternatively:
\begin{equation}\left[  \square - \frac{1}{4} 
    \left(e_\beta\rfloor\square\vartheta^\beta \right) \right]
  \eta_\alpha = \ell^2\,\Sigma\kern-1.1em\nearrow_\alpha\;.
  \label{eq:decomp-second'}
\end{equation}

\subsection{Yilmaz-Rosen and Schwarzschild\\ solution compared}
\label{sec:YR-sol} 
The Yilmaz-Rosen metric and the corresponding orthonormal coframe were
displayed in \eqref{YILMAZ} to \eqref{COFRAME}. In order to compare
the Yilmaz-Rosen metric with the Schwarzschild metric, we transform
the former one from the isotropic coordinates used in \eqref{YILMAZ}
into Schwarzschild coordinates and Taylor expand it:
\begin{subequations}
\begin{align}
g^{\text{YR}}_{00} & = 1-\frac{2m}{r} \qquad \quad \,+ \frac{25}{6}
\frac{m^3}{r^3} + \mathcal{O} 
\left( \tfrac{m^4}{r^4}\right) \;,\\
g^{\text{YR}}_{11} & = 1+ \frac{2m}{r} + \frac{5m^2}{r^2} +
\frac{9m^3}{r^3}\; +\, \mathcal{O}\left( \tfrac{m^4}{r^4} \right) \;.
\end{align}
\end{subequations}
For the Schwarzschild metric in Schwarzschild coordinates we find:
\begin{subequations}
\begin{align}
  g^{\text{SS}}_{00} & = 1-\frac{2m}{r} \qquad \text{(exact)}\;,\\ 
  g^{\text{SS}}_{11} & = 1+ \frac{2m}{r} + \frac{4m^2}{r^2} +
  \frac{8m^3}{r^3}\, +\, \mathcal{O}\left( \tfrac{m^4}{r^4} \right) \;.
\end{align}
\end{subequations}

Their $g_{00}$ components are equal up to second order. The radial
components $g_{11}$ begin to differ slightly in the second order.
Therefore the Yilmaz-Rosen solution is consistent with the classical
tests of general relativity and can, in particular, describe the
post-Newtonian perihelion advance correctly, see Synge \cite[page 296,
footnote 1]{synge71}. It requires further investigations to decide
whether the two solutions can be observationally distinguished by
strong gravity effects in close binary pulsar systems.

\subsection{Yilmaz-Rosen metric motivated}\label{YR-mot}
A viable theory of gravitation should be consistent with the
\emph{local} equivalence principle. Let us consider an electromagnetic
wave of frequency \(\omega\) in the gravitational field of a point
mass $m$. The frequency shift due to the propagation from a point with
radial coordinate $r$ to one with $r+\Delta r$ reads ($c=G=1$):
\begin{equation} 
\frac{\Delta \omega}{\omega} = \Delta U = -\frac{m\Delta r}{r^2} \;.
\label{eq:equiv-local} \end{equation}

According to Mashhoon \cite{mash97}, Yilmaz effectively proposed to
extend the local equivalence principle to a \emph{global} one.  With
this idea in mind, we can tentatively integrate \eqref{eq:equiv-local}:
\begin{equation} \int_{r_0}^{r_1} \frac{d\omega}{\omega} = 
  \int_{r_0}^{r_1} dU \;.\end{equation} We solve the integrals and
find
\begin{equation}\frac{\omega(r_1)}{\omega(r_0)} = \exp[U(r_1)-U(r_0)]\;.
\end{equation} Using the Newtonian potential explicitly, 
\(U=-\frac{m}{r}\), and taking the limit \(r_1\rightarrow \infty\),
yields
\begin{equation} \frac{\omega(\infty)}{\omega(r)}= \exp\left(\frac{m}{r}
  \right)\;, \quad \text{and thus} \qquad \frac{\Delta
    t(\infty)}{\Delta t(r)} = \exp \left( - \frac{m}{r} \right) \;.
\end{equation} Since \(\Delta t(\infty)\) is not influenced by
gravity, one can directly read off \begin{equation}
  \vartheta^{\hat{t}} = e^{-\frac{m}{r}} \,dt \quad \text{or} \qquad
  {g}_{00} = e^{-\frac{2m}{r}}\;.\end{equation}

Following the pattern of the components of the Schwarzschild metric,
we now define \(g_{11}\) as the inverse of \(g_{00}\):
\begin{equation}\label{YR-Schwarz} \widetilde{g} = 
  e^{-\frac{2m}{r}}dt^2-e^{\frac{2m}{r}}dr^2-r^2\left(d\theta^2+
    \sin^2\theta\,d\varphi^2\right)\;.  \end{equation} However, the
Taylor approximation of this metric (which is \emph{not} the
Yilmaz-Rosen metric of \eqref{YILMAZ}) reads:
\begin{subequations}
\begin{align}
\widetilde{g}_{00} & = 1-\frac{2m}{r} + \frac{2m^2}{r^2} - \frac{4}{3}
\frac{m^3}{r^3} + \mathcal{O} 
\left( \tfrac{m^4}{r^4}\right) \;,\\
\widetilde{g}_{11} & = 1+ \frac{2m}{r} + \frac{2m^2}{r^2} +
\frac{4}{3}\frac{m^3}{r^3} + \mathcal{O}\left( \tfrac{m^4}{r^4}
\right) \;. 
\end{align}
\end{subequations}
The $\widetilde{g}_{00}$ component differs from $g^{\text{SS}}_{00}$
already to second order. And the deviation of $\widetilde{g}_{11}$
from $g^{\text{SS}}_{11}$ is doubled in comparison to that of the
Yilmaz-Rosen $g^{\text{YR}}_{11}$.  Therefore, in order to approximate
the experimentally well verified Schwarzschild metric in an optimal
way, we choose the forefactors of \eqref{YR-Schwarz} as metric
components in \emph{isotropic} coordinates, which eventually leads to
the Yilmaz-Rosen metric:
\begin{equation} g^{\text{YR}}=e^{-\frac{2m}{r}}\,dt^2-e^{\frac{2m}{r}}
  \,\left(dx^2 +dy^2 +dz^2\right)\;.\end{equation}

\subsection{Yilmaz-Rosen solution and the vacuum field equation}
The Yilmaz-Rosen metric \eqref{YILMAZ}, keeping in mind
\eqref{ORTHONORMAL} and \eqref{COFRAME}, fulfills the tracefree field
equation \eqref{eq:trace-decomp} [or, alternatively,
\eqref{eq:decomp-second} or \eqref{eq:decomp-second'}] with vanishing
source. For a proof compare the corresponding computer algebra program
in Sec.\ \ref{sec:ca}.
As we saw, the Yilmaz-Rosen solution is consistent with the classical
tests of general relativity. However, its integration constant $m$
cannot be directly identified with the source of a spherical body.

For that reason, we take recourse to the trace equation
\eqref{eq:trace}. In the vacuum case we find
\begin{equation}\lambda(x)=\frac{1}{4}{}\,^\star\left( 
    \vartheta^\gamma\wedge\square\eta_\gamma\right)\,.
\end{equation} The right hand side of this equation can be easily 
calculated with our computer algebra program, see also \cite{kani97}.
Therefore the Yilmaz-Rosen metric solves the vacuum field equation
\eqref{KI-fieldeq} provided the `cosmological' function is prescribed
as follows:
\begin{equation}\lambda(x)=-\left( \frac{m}{r^2} e^{-\frac{m}{r}} 
  \right) ^2\,.
\end{equation}

\begin{figure}
\includegraphics[width=\textwidth]{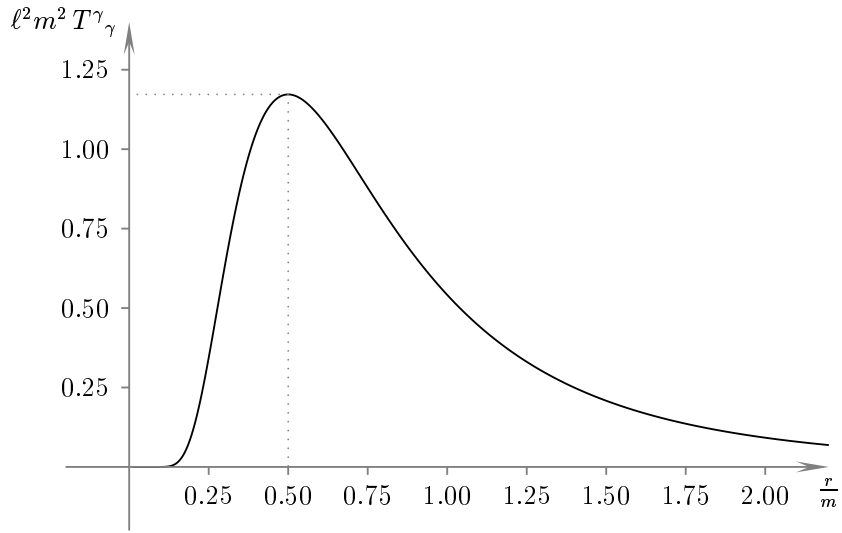}
\caption[]{\emph{The prescribed matter distribution}
  $\ell^2\,T^\gamma{}_\gamma = \left( \frac{2m}{r^2}
    e^{-\frac{m}{r}}\right) ^2$. \emph{Most of the matter is inside
  the Schwarzschild radius} $r_{s}=2m$.}\label{fig:matter}
\end{figure}

Since such an ad hoc structure looks too implausible to us, we can
change horses at this moment: We reject the `cosmological' function of
Kaniel-Itin and put it to zero. Then we can mimic this function
$\lambda(x)$ by means of the energy-momentum trace
$-\ell^2\,T^\gamma{}_\gamma/4$, cf.\ \eqref{eq:trace} and
\eqref{eq:energy-tensor}:
\begin{equation} T^\gamma{}_\gamma =
    \left( \frac{2m}{\ell r^2} e^{-\frac{m}{r}}\right) ^2 \;.
\label{eq:matter-distrib}\end{equation} 
This energy-momentum trace is plotted in Fig.\ \ref{fig:matter}.
Therefore, the Yilmaz-Rosen solution fulfills the field equations
\begin{equation} \square \eta_\alpha - \frac{1}{4} e_\alpha
\rfloor \left( \vartheta^\gamma \wedge \square \eta_\gamma \right)
=0\qquad\text{and}\qquad \ensuremath{{}^\star} \left( \vartheta^\gamma
\wedge \square \eta_\gamma \right) = -\left( \frac{2m}{\ell r^2}
e^{-\frac{m}{r}}\right) ^2 \;.\end{equation} 

Taking an ideal fluid for the description of matter, then, for
vanishing pressure, \(p=0\), we have \(T^\gamma{}_\gamma=\rho\).
Therefore we can understand the above computation as a matter
distribution \eqref{eq:matter-distrib} which can be viewed as
(probably unphysical) star model. The matter of such a star reaches to
infinity, but it decreases exponentially. We find the
maximum of the distribution at \(r=\kern.1em%
\raise.5ex\hbox{\scriptsize \ensuremath{m}}%
\kern-.1em/\kern-.15em%
\lower.25ex\hbox{\scriptsize \ensuremath{2}}\), compare with Fig.\ 
\ref{fig:matter}. Most of the star mass is concentrated inside the
Schwarzschild radius \(r_s=2m\). The volume integral over
\(T^\gamma{}_\gamma\) yields the total mass \(m\) of the
star:
\begin{multline}
  \int T^\gamma{}_\gamma\, dV=4\pi \int\limits_0^{\infty}
  T^\gamma{}_\gamma\, r^2\, dr = \int\limits_0^\infty \frac{16\pi
    m^2}{\ell^2 r^2}\,e^{-\frac{2m}{r}} \, dr\;
  \\\stackrel{x:=-\frac{2m}{r},\; dr=\frac{2m}{x^2}\,dx}{=}\;
  \frac{8\pi m}{\ell^2} \int\limits_{-\infty}^0 e^x \, dx = \frac{8\pi
    m}{\ell^2}=M\;,
\end{multline} where $M$ is the mass in conventional units.

As a result, we can interpret the constant \(m\) of the Yilmaz-Rosen 
metric as the mass of a star, but this mass
is distributed in a probably unphysical way. 

\section{Conclusion}

We posed four questions about the Kaniel-Itin model in Sec.\ 1. We
will try to answer them in turn:
\begin{itemize}
\item[(i)] We can put the energy-momentum 3-form of matter on the
  right hand side of the KI-field equation \eqref{WAVE}, see
  \eqref{KI-fieldeq}:
\begin{equation}\label{KI-fieldeq'}
  \left[\square +\lambda(x)\right]\eta_\alpha=\ell^2\,
  \frac{\delta
    L_{\text{mat}}}{\delta\vartheta^\alpha}=:\ell^2\,\Sigma_\alpha\,.
\end{equation}
\item[(ii)] The Yilmaz-Rosen metric can be accommodated to
  \eqref{KI-fieldeq'} in the following sense: We decompose
  \eqref{KI-fieldeq'} in vacuum into
\begin{subequations}
\begin{align} -\square \eta_\alpha + \frac{1}{4} e_\alpha \rfloor
  \left( \vartheta^\beta \wedge \square \eta_\beta \right) &=0
  \;,\label{eq:trace-decomp'} \\ \frac{1}{4}\,
  ^\star\left(\vartheta^\beta \wedge \square \eta_\beta\right) &=
  \lambda(x)\;.
\label{eq:trace'}\end{align}
\end{subequations}
These equations are fulfilled by the Yilmaz-Rosen metric, provided we
prescribe the `cosmological' function $\lambda(x)$ in the following
way:
\begin{equation}\label{lambdaofx}
  \lambda(x)= \frac{1}{4}\, ^\star\left(\vartheta^\beta \wedge \square
    \eta_\beta\right) =-\left( \frac{m}{r^2} e^{-\frac{m}{r}}\right)^2
  \;.\end{equation} If one used a wave equation as field equation,
i.e., if in \eqref{KI-fieldeq'} one put $\lambda(x)=0$, then one could
find the Yilmaz-Rosen solution for the matter distribution
$T^\gamma{}_\gamma=\left( \frac{2m}{\ell r^2} e^{-\frac{m}{r}}\right)^2$.
\item[(iii)] There doesn't exist a consistent variational
  principle for arriving at \eqref{KI-fieldeq'}. Maybe one is able to
  find one for the tracefree vacuum equation \eqref{eq:trace-decomp'}.
\item[(iv)] The constraints of Kaniel-Itin on the variations amount
  to getting rid of the independence of the variations of the metric,
  provided the variations of the coframe are prescribed.
\end{itemize}
Is the model of Kaniel \& Itin viable? Well, it is presently in
intensive care~\dots And it is a beautiful model anyways.

\paragraph{Acknowledgments:} We are grateful to Shmuel Kaniel and
Yakov Itin for interesting discussions and most helpful remarks. We
thank Bahram Mashhoon, Eckehard Mielke, and Yuri Obukhov for useful
comments and hints.  

\appendix
\section{Hodge star and $\eta$-basis}
\subsection{Elementary relations for the star etc.}\label{formulae}

We now collect some rules for calculations with the Hodge star, where
\(\psi\) and \(\phi\) are forms of the \emph{same} degree \(p\) (see
\cite{hehl95}):
\begin{gather}
  \ensuremath{{}^\star} \ensuremath{{}^\star} \psi =
  (-1)^{p(n-p)+\text{ind}(g)} \, \psi
\label{eq:hodge-involutiv} \;, \\
\ensuremath{{}^\star} \psi \wedge \phi = \ensuremath{{}^\star} \phi
\wedge \psi
\label{eq:hodge-wedge}\;.
\end{gather}
The \emph{index} \(\text{ind}(g)\) of a metric is the number of minus
signs if it is in diagonal form.  Furthermore one has the useful rules
\begin{subequations}
\label{eq:hodge-coframe}
\begin{align}
  e_\alpha \rfloor \ensuremath{{}^\star} \psi
  \label{eq:hodge-coframe1} & = \ensuremath{{}^\star} \left(
    \psi\wedge \vartheta_\alpha \right) \;.\\ e_\alpha \rfloor \psi
  \label{eq:hodge-coframe2} & = (-1)^{\text{ind}(g)} \;
  \ensuremath{{}^\star} \!\left( \vartheta_\alpha \wedge
    \ensuremath{{}^\star} \psi\right) \;, \\ \ensuremath{{}^\star}
  \left( e_\alpha \rfloor \psi \right) & = (-1)^{p-1} \;
  \vartheta_\alpha\wedge \ensuremath{{}^\star} \psi \;,
  \label{eq:hodge-coframe3}\\ \ensuremath{{}^\star}\left( e_\alpha
    \rfloor \ensuremath{{}^\star} \psi \right) &=
  (-1)^{(p+1)+\text{ind}(g)} \, \psi \wedge \vartheta_\alpha \;.
  \label{eq:hodge-coframe4}
\end{align}
\end{subequations}
Sometimes we also need the formula:
\begin{equation} \vartheta^\mu \wedge 
  \left( e_\mu \rfloor \psi\right) = p\,\psi\;.\label{eq:expand}
\end{equation}

With these rules one can determine the \(\eta\)-basis, cf.\ 
\eqref{etabasis}:
\begin{subequations}
\label{eq:eta-def-all}
\begin{align}
  \eta := \ensuremath{{}^\star} 1 & = \frac{1}{n!}
  \eta_{\alpha_1\dots\alpha_n} \vartheta^{\alpha_1} \wedge \dots
  \wedge \vartheta^{\alpha_n} = \frac{1}{n!} \sqrt{|\det
    g_{\mu\nu}|}\, \vartheta^{\alpha_1} \wedge \dots \wedge
  \vartheta^{\alpha_n} \;,\\ \eta^{\alpha_1\dots\alpha_p} & :=
  \ensuremath{{}^\star} \left(\vartheta^{\alpha_1} \wedge \dots\wedge
    \vartheta^{\alpha_p}\right) = \frac{1}{(n-p)!}
  \eta^{\alpha_1\dots\alpha_p}{}_{\alpha_{p+1}\dots\alpha_n}
  \vartheta^{\alpha_{p+1}} \wedge \dots \wedge \vartheta^{\alpha_n}
  \notag \\ & = \frac{\sqrt{|\det g_{\mu\nu}|}}{(n-p)!} \,
  g^{\alpha_1\beta_1}\cdots g^{\alpha_p\beta_p}\,
  \epsilon_{\beta_1\cdots\beta_p\alpha_{p+1}\cdots \alpha_n}
  \vartheta^{\alpha_{p+1}} \wedge \dots \wedge \vartheta^{\alpha_n}
\label{eq:eta-def} \;, \\  
\eta^{\alpha_1\dots\alpha_n} &:= \ensuremath{{}^\star}
\!\left(\vartheta^{\alpha_1} \wedge \dots\wedge
  \vartheta^{\alpha_n}\right) \notag\\ & = \sqrt{|\det g_{\mu\nu}|}\,
g^{\alpha_1\beta_1}\cdots g^{\alpha_n\beta_n}\,
\epsilon_{\beta_1\cdots\beta_n} = \frac{1}{\sqrt{|\det g_{\mu\nu}|}}
\;\epsilon^{\alpha_1\cdots\alpha_n}
\label{eq:eta-funktion}\;. 
\end{align}
\end{subequations}

Two helpful rules, which connect the different elements of the
\(\eta\)-basis, read
\begin{subequations}
\begin{gather} \eta^{\alpha_1\dots \alpha_p}{}_\mu = e_\mu \rfloor
  \eta^{\alpha_1\dots \alpha_p} \;, \\ \vartheta^\mu \wedge
  \eta^{\alpha_1\dots \alpha_p} = \sum_{i=1}^p (-1)^{p-i}
  g^{\mu\alpha_i} \eta^{\alpha_1\dots\alpha_{i-1}\alpha_{i+1}
    \dots\alpha_p}\;. \label{eq:rettung}\end{gather}
\end{subequations}

In case of independent variations of the metric components
\(g_{\alpha\beta}\), we need the rules 
\begin{align} \delta g^{\alpha\beta} & = - g^{\alpha\gamma} g^{\delta\beta}
\,\delta g_{\gamma\delta} \;, \label{eq:var-metrikinv}\\
\label{eq:var-metrikdet} \delta
\left[\det\left(g_{\mu\nu}\right)\right] & = \det\left(g_{\mu\nu}\right)
g^{\alpha\beta} \, \delta g_{\alpha\beta}\;. \end{align}

\subsection{Variation of the $\eta$-basis}
With these definition and rules, we can compute\label{sec:2.2} 
a general variation of the \(\eta\)-basis, involving the fields
\(\vartheta^\alpha\) and \(g_{\alpha\beta}\):
\begin{equation}\begin{split}
  \delta \eta^{\beta_1\dots\beta_p} & = \frac{1}{(n-p)!} \,\delta
  \left( \eta^{\beta_1\dots\beta_p}{}_{\beta_{p+1}\dots\beta_n}
    \vartheta^{\beta_{p+1}}\wedge \dots \wedge \vartheta^{\beta_n}
  \right) \\ &
  \stackrel{\text{\makebox[0pt]{\ensuremath{\eqref{eq:eta-def}}}}}{=}
  \quad \frac{1}{(n-p)!}\,
  \eta^{\beta_1\dots\beta_p}{}_{\beta_{p+1}\dots\beta_n} \delta \left(
    \vartheta^{\beta_{p+1}}\wedge \dots \wedge \vartheta^{\beta_n}
  \right) + \frac{1}{(n-p)!}\\ &\;\: \times \delta \left( \sqrt{|\det
      g_{\mu\nu}|}\, g^{\alpha_1\beta_1}\cdots g^{\alpha_p\beta_p}\,
    \epsilon_{\alpha_1\cdots\alpha_p\beta_{p+1}\cdots
      \beta_n}\right)\, \vartheta^{\beta_{p+1}} \wedge \dots \wedge
  \vartheta^{\beta_n} \\ &
  \stackrel{\text{\makebox[0pt]{\ensuremath{\substack{\eqref{eq:var-leibniz},
            \eqref{eq:var-metrikdet},\\ \eqref{eq:var-metrikinv},
            \eqref{eq:eta-def}}}}}}{=} \qquad \frac{n-p}{(n-p)!}\,
  \eta^{\beta_1\dots\beta_p}{}_{\beta_{p+1}}{}_{\beta_{p+2}\dots\beta_n}
  \left(\delta\vartheta^{\beta_{p+1}}\right) \wedge \left(
    \vartheta^{\beta_{p+2}}\wedge \dots \wedge \vartheta^{\beta_n}
  \right) \\ & \qquad \qquad \qquad + \left( \frac{1}{2}
    g^{\kappa\lambda} \eta^{\beta_1\dots \beta_p} - \sum_{i=1}^p
    g^{\kappa\beta_i} \eta^{\beta_1\dots \beta_{i-1}\lambda
      \beta_{i+1} \dots\beta_p} \right)\, \delta g_{\kappa\lambda} \\ 
  &
  \stackrel{\text{\makebox[0pt]{\ensuremath{\eqref{eq:eta-def}}}}}{=}
  \quad \delta\vartheta^{\beta_{p+1}} \wedge
  \eta^{\beta_1\dots\beta_p}{}_{\beta_{p+1}} + \left( \frac{1}{2} \,
    g^{\kappa\lambda} \eta^{\beta_1\dots \beta_p} \right. \\ & \qquad
  \qquad \qquad \qquad \left.- \sum_{i=1}^p (-1)^{p-i}
    g^{\kappa\beta_i} g^{\lambda\rho} e_\rho \rfloor
    \eta^{\beta_1\dots \beta_{i-1}\beta_{i+1} \dots\beta_p} \right)\,
  \delta g_{\kappa\lambda} \\ &
  \stackrel{\text{\makebox[0pt]{\ensuremath{\eqref{eq:rettung}}}}}{=}
  \quad \delta\vartheta^{\beta_{p+1}} \wedge
  \eta^{\beta_1\dots\beta_p}{}_{\beta_{p+1}} \\ & \qquad \qquad +
  \left( \frac{1}{2} g^{\kappa\lambda} \eta^{\beta_1\dots \beta_p} -
    g^{\lambda\rho} e_\rho \rfloor \left(\vartheta^\kappa \wedge
      \eta^{\beta_1 \dots \beta_p} \right) \right)\, \delta
  g_{\kappa\lambda}\;.
\end{split}\end{equation}
Hence
\begin{equation}
\label{eq:var-eta'} \delta \eta^{\beta_1\dots\beta_p} =
\delta\vartheta^\mu \wedge \left( e_\mu \rfloor
  \eta^{\beta_1\dots\beta_p} \right) + \left( \vartheta^{(\kappa|}
  \wedge \eta^{\beta_1 \dots \beta_p|\lambda)} - \frac{1}{2} \,
  g^{\kappa\lambda} \eta^{\beta_1\dots\beta_p} \right) \, \delta
g_{\kappa\lambda}\;.
\end{equation}

\subsection{Deduction of \eqref{eq:var-hodge}}\label{deduc33}
We start with \eqref{varphi}. We abbreviate \(\psi_p:=\ptetr\) and
get, using the variation \eqref{eq:var-eta'} of
\(\eta^{\beta_1\dots\beta_p}\):
\begin{equation}\begin{split}
    \psi_p \wedge \delta \ensuremath{{}^\star} \phi &= \delta \phi
    \wedge \eta^{\beta_1\dots\beta_p} - \delta \psi_p \wedge
    \ensuremath{{}^\star} \phi + \phi \wedge \delta \vartheta^\mu
    \wedge \eta^{\beta_1\dots\beta_p}{}_{\mu} \\ & \qquad + \phi
    \wedge \left( \vartheta^\kappa \wedge \eta^{\beta_1 \dots
        \beta_p\lambda} - \frac{1}{2} \, g^{\kappa\lambda}
      \eta^{\beta_1\dots\beta_p} \right) \, \delta g_{\kappa\lambda}
    \\ & \stackrel{\text{\makebox[0pt]{\ensuremath{\eqref{eq:eta-def},
            \eqref{eq:hodge-wedge}}}}}{=} \qquad \psi_p\wedge
    \ensuremath{{}^\star} \left(\delta \phi\right) - \delta \psi_p
    \wedge \ensuremath{{}^\star} \phi + \phi\wedge \delta\vartheta^\mu
    \wedge \left( e_\mu \rfloor \eta^{\beta_1\dots\beta_p}\right) \\ &
    \qquad \qquad + \left(\phi \wedge \vartheta^\kappa \wedge \left(
        g^{\lambda\rho} e_\rho \rfloor
        \eta^{\beta_1\dots\beta_p}\right) - \frac{1}{2} \,\psi_p
      \wedge g^{\kappa\lambda}\; \ensuremath{{}^\star} \phi \right) \,
    \delta g_{\kappa\lambda} \\ & =\psi_p\wedge \left(
      \ensuremath{{}^\star} \left(\delta \phi\right) - \frac{1}{2}\,
      \ensuremath{{}^\star} \phi \; g^{\kappa\lambda}\, \delta
      g_{\kappa\lambda} \right) - \delta\vartheta^\mu \wedge \left(
      e_\mu \rfloor \psi_p\right) \wedge \ensuremath{{}^\star} \phi \\ 
    & \quad + (-1)^p \left( \, \delta\vartheta^\mu \wedge \phi \wedge
      \left(e_\mu \rfloor \ensuremath{{}^\star} \psi_p \right) + \,
      g^{\lambda\rho} e_\rho \rfloor \left( \phi \wedge
        \vartheta^\kappa \right) \wedge \ensuremath{{}^\star} \psi_p
      \; \delta g_{\kappa\lambda}\right) \\ & = \psi_p\wedge \left(
      \ensuremath{{}^\star} \left(\delta \phi\right) - \frac{1}{2}\,
      \ensuremath{{}^\star} \phi \; g^{\kappa\lambda}\, \delta
      g_{\kappa\lambda} \right) + \delta\vartheta^\mu \wedge \Big(
    e_\mu \rfloor \left( \phi \wedge \ensuremath{{}^\star} \psi_p
    \right) \\ &\qquad \quad - \left( e_\mu \rfloor \phi \right)
    \wedge \ensuremath{{}^\star} \psi_p - e_\mu \rfloor \left( \psi_p
      \wedge \ensuremath{{}^\star} \phi \right) + (-1)^p \psi_p \wedge
    \left( e_\mu \rfloor \ensuremath{{}^\star}\phi\right)\Big) \\ 
    &\qquad \qquad \qquad \qquad \qquad \qquad + \psi_p \wedge
    \vartheta^\lambda \wedge \ensuremath{{}^\star} \left( \phi \wedge
      \vartheta^\kappa \right) \, \delta g_{\kappa\lambda} \\ &
    \stackrel{\text{\makebox[0pt]{\ensuremath{\eqref{eq:hodge-wedge}}}}}{=}
    \quad \psi_p\wedge \ensuremath{{}^\star} \left(\delta \phi\right)
    - \psi_p \wedge \ensuremath{{}^\star} \left( \delta \vartheta^\mu
      \wedge \left( e_\mu \rfloor \phi \right) \right) + \psi_p \wedge
    \delta \vartheta^\mu \wedge \left( e_\mu \rfloor
      \ensuremath{{}^\star} \phi\right)\\ &\qquad \qquad - \psi_p
    \wedge \frac{1}{2}\, \ensuremath{{}^\star} \phi \;
    g^{\kappa\lambda}\, \delta g_{\kappa\lambda} + \psi_p\wedge
    \vartheta^\lambda \wedge g^{\kappa\rho} e_\rho \rfloor \left(
      \ensuremath{{}^\star} \phi \right) \, \delta g_{\kappa\lambda}
    \;.
\end{split}\end{equation}

Since \(\psi_p:=\ptetr\) is constructed with \(p\) arbitrary
\(\vartheta^\alpha\)'s, we conclude for an arbitrary \(p\)-form \(\phi\):
\begin{equation}
\begin{split} \left(\delta \ensuremath{{}^\star}
        -\ensuremath{{}^\star} \delta\right) \phi & = \;\delta
      \vartheta^\alpha \wedge \left(e_\alpha \rfloor
        \ensuremath{{}^\star} \phi \right) - \ensuremath{{}^\star}
      \big[ \delta \vartheta^\alpha \wedge \left( e_\alpha \rfloor
        \phi \right)\big] \label{eq:var-hodge'} \\ & + \delta
      g_{\alpha\beta}\Big[\vartheta^{(\alpha}\! \wedge\!
      ( e^{\beta)}
 \rfloor \ensuremath{{}^\star} \phi ) -
      \frac{1}{2}\,g^{\alpha\beta}\;
      \ensuremath{{}^\star} \phi\Big] \;.
\end{split}
\end{equation}

For the special choice of an \emph{orthonormal} coframe, we have \(\delta
g_{\alpha\beta}=0\). In this case the two last
summands vanish:
\begin{equation}
  \left(\delta \ensuremath{{}^\star} -\ensuremath{{}^\star}
    \delta\right) \phi = \;\delta \vartheta^\alpha \wedge
  \left(e_\alpha \rfloor \ensuremath{{}^\star} \phi \right)\; -\;
  \ensuremath{{}^\star} \Big[ \delta \vartheta^\alpha \wedge \left(
    e_\alpha \rfloor \phi \right)\Big]\;. \label{eq:var-hodge-ortho}
\end{equation}

\section{Computer algebra program}\label{sec:ca}
The following \texttt{Reduce} program was written with the help of the
\texttt{Excalc} package, see \cite{excalc87,reduce93}.\footnote{This
  program works properly only with a new patch of \texttt{Excalc},
  fixing an earlier bug on the hodge dual of scalars. Older versions
  of \texttt{Excalc} need as additional input the following function
  which should be put in after \texttt{Excalc} has been loaded:\\ 
  \texttt{ symbolic procedure dual0 u; \\ $\qquad$ (multpfsq(mkwedge
    ('wedge . basisforml!*),\\ $\qquad\qquad$ simpexpt
    list(mk!*sq(absf!* numr x ./\\ $\qquad\qquad\qquad$ absf!* denr
    x),'(quotient 1 2))))\\ $\qquad$ where x = simp!* detm!*;\\ }} It
verifies (i) the decompositon (\ref{eq:base-rumpf}) of the
KI-Lagrangian, (ii) that the Yilmaz-Rosen metric fulfills the
tracefree KI-vacuum field equation (74a) and (77), and (iii) the
validity of Eq.\ (89).
\begin{verbatim}
% file kaniti.exi, 1998-01-11, fwh+fg     %in "kaniti.exi";

load_package excalc$

%
% Basic definitions:
%

pform psi=0, r=0, lam=0$
fdomain psi=psi(x,y,z), r=r(x,y,z), lam=lam(x,y,z)$

coframe o(t) =  psi    * d t  + a*sin(x)* d x,
        o(x) = (1/psi) * d x  + b*sinh(z)*d y,
        o(y) = (1/psi) * d y,
        o(z) = (1/psi) * d z  with signature(+1,-1,-1,-1)$

frame e$
sgn :=-1$

%
% Checking the decompositon (58) of the KI-Lagrangian
%

pform v0rumpf4=4, v1rumpf4=4, v2rumpf4=4, v3rumpf4=4, 
      v4rumpf4=4, vki4=4$

v0rumpf4 :=    o(a)  ^ #   o(-a)/4$
v1rumpf4 :=  d o(a)  ^ # d o(-a)$
v2rumpf4 := (d o(-a) ^     o(a)) ^ # (d o(-b) ^ o(b) )$
v3rumpf4 := (d o(a)  ^     o(b)) ^ # (d o(-a) ^ o(-b))$
v4rumpf4 := (d o(-a) ^     o(b)) ^ # (d o(-b) ^ o(a) )$

vki4 := rho0 * v0rumpf4 + rho1 * v1rumpf4 + rho2 * v2rumpf4
                        + rho3 * v3rumpf4 + rho4 * v4rumpf4;

rho0 := 0; rho1 := 2; rho2 := 0; rho3 := 0; rho4 :=-1; 

diff := vki4 -  (  d    o(a)   ^ #(   d   o(-a)) 
             +   #(d (# o(a))) ^ #(# (d # o(-a))) );
             
diff := diff;

% diff has to vanish. Note that our check is not for 
% a general coframe. In the following we choose the 
% Yilmaz-Rosen coframe and set a=0, b=0:

a:=0$
b:=0$

r**2 := (x**2+y**2+z**2)$
@(r,x):= x/r; @(r,y):= y/r; @(r,z):= z/r$ 
psi := exp(-m/r)$ 

pform dalembertcof1(a)=1, dalemberteta3(a)=3, 
kifeqtrfreea3(a)=3, kifeqtrfreeb3(a)=3$

dalembertcof1(a):= -d(#(d(#   o(a) ))) - #(d(#(d  o(a) )));
dalemberteta3(a):= -d(#(d(#(# o(a))))) - #(d(#(d(#o(a)))));

%
% Checking the vacuum field equation (1) of Kaniel & Itin
%

% lhs of Eq.(74a)
kifeqtrfreea3(a) := dalemberteta3(a) - e(a) _| 
                    ( o(b) ^ dalemberteta3(-b) )/4;

% lhs of Eq.(77)
kifeqtrfreeb3(a) := dalemberteta3(a) - ( e(-b) _| 
                    (dalembertcof1(b))/4 )  ^  # o(a); 

% Eq.(88)
lam  := #(o(b) ^ dalemberteta3(-b))/4;

% Eq.(89)
lam - ( - ((m/r**2)*e**(-m/r))**2);

% Eq.(90)
energytrace := - 4 * lam / ell**2;

end$

\end{verbatim}

\newpage
\noindent
\end{document}